\documentclass[epjc3,final]{mysvjour3}
\pdfoutput=1
\usepackage{amsmath,amssymb,amsfonts}
\usepackage{hyperref}
\usepackage{graphicx}
\usepackage{color}
\usepackage{subfigure}
\usepackage{cite}
\usepackage{listings}

\definecolor{shaded}{RGB}{210,210,210}
\hypersetup{%
   pdfborder={0 0 0},
   pdftitle={GoSam 2.0},
   pdfauthor={G. Cullen, H. van Deurzen, N. Greiner, G. Heinrich, G. Luisoni, %
              P. Mastrolia, E. Mirabella, G. Ossola, T. Peraro, J. Schlenk, J.F.von Soden-Fraunhofen, F. Tramontano},
   pdfsubject={GoSam},
   pdfkeywords={NLO calculations, automation, hadron colliders %
          PACS 12.38.-t, 12.38.Bx, 12.60.-i},
   pdflang={en_US}
}

\graphicspath{{pics/}}

\newcommand{\GOSAM}{{\textsc{Go\-Sam}}}
\newcommand{\gosamv}{{\textsc{Go\-Sam}}{-2.0}}

\newcommand{\QGRAF}{{\texttt{QGRAF}}}
\newcommand{\FORM}{{\texttt{FORM}}}
\newcommand{\SPINNEY}{{\texttt{spin\-ney}}}
\newcommand{\HAGGIES}{{\texttt{hag\-gies}}}
\newcommand{\SAMURAI}{{\textsc{Sa\-mu\-rai}}}
\newcommand{\GOLEMVC}{{\textsc{Go\-lem95C}}}
\newcommand{\NINJA}{{\textsc{Ninja}}}

\newcommand{\FORTRAN}{{\texttt{For\-tran}}}
\newcommand{\python}{{\texttt{Py\-thon}}}

\newcommand{\tHV}{{'t\,Hooft Veltman}}

\newcommand{\mrm}[1]{\mathrm{#1}}

\newenvironment{kinematics}{%
\fontsize{8pt}{10pt}\selectfont%
\begin{tabular}{l|rrrr}
   & \multicolumn{1}{c}{$E$}
   & \multicolumn{1}{c}{$p_x$} 
   & \multicolumn{1}{c}{$p_y$}
   & \multicolumn{1}{c}{$p_z$} \\
\hline}{%
\end{tabular}}

\newcommand{\bea}{\begin{eqnarray*}}
\newcommand{\eea}{\end{eqnarray*}\noindent}
\newcommand{\be}{\begin{equation}}
\newcommand{\ee}{\end{equation}\noindent}
\newcommand{\bcen}{\begin{center}}
\newcommand{\ecen}{\end{center}}
\newcommand{\bfl}{\begin{flushleft}}
\newcommand{\efl}{\end{flushleft}}
\newcommand{\nn}{\nonumber}
\def\eps{\epsilon}
\def\ket#1{|{#1}\rangle}
\def\bra#1{\langle{#1}|}

\newcommand{\diff}[1][{}]{{\mathrm{d}}^{#1}\!}
\newcommand{\calst}{\mbox{$\cal S$}}

\journalname{Eur. Phys. J. C}

\begin{document}

\title{\GOSAM{}-2.0: a tool for automated one-loop calculations within the Standard Model and beyond}
\author{%
       Gavin~Cullen\thanksref{desy} 
  \and Hans van Deurzen\thanksref{vandeurzen,mpp}
  \and Nicolas~Greiner\thanksref{greiner,mpp}
  \and Gudrun~Heinrich\thanksref{heinrich,mpp}
  \and Gionata~Luisoni\thanksref{luisoni,mpp}
  \and Pierpaolo~Mastrolia\thanksref{mastrolia,mpp,padova}
  \and Edoardo Mirabella\thanksref{mirabell,mpp}
  \and Giovanni~Ossola\thanksref{ossola,ny,ny2}
  \and Tiziano Peraro\thanksref{peraro,mpp}
  \and Johannes Schlenk\thanksref{jschlenk,mpp}
  \and Johann Felix von Soden-Fraunhofen\thanksref{jfsoden,mpp}
  \and Francesco~Tramontano\thanksref{tramontano,napoli}
}

\thankstext{vandeurzen}{\email{hdeurzen@mpp.mpg.de}}
\thankstext{greiner}{\email{greiner@mpp.mpg.de}}
\thankstext{heinrich}{\email{gudrun@mpp.mpg.de}}
\thankstext{luisoni}{\email{luisonig@mpp.mpg.de}}
\thankstext{mastrolia}{\email{Pierpaolo.Mastrolia@cern.ch}}
\thankstext{mirabell}{\email{mirabell@mpp.mpg.de}}
\thankstext{ossola}{\email{gossola@citytech.cuny.edu}}
\thankstext{peraro}{\email{peraro@mpp.mpg.de}}
\thankstext{jschlenk}{\email{jschlenk@mpp.mpg.de}}
\thankstext{jfsoden}{\email{jfsoden@mpp.mpg.de}}
\thankstext{tramontano}{\email{Francesco.Tramontano@cern.ch}}

\institute{%
\parindent=0pt %
     Deutsches Elektronen-Synchrotron DESY, Zeuthen, Germany\label{desy}
\and Max-Planck-Institut f\"ur Physik, M\"unchen, Germany\label{mpp}
\and Dipartimento di Fisica, Universit\`a di Padova, Italy\label{padova}
\and New York City College of Technology, City University of New York, USA\label{ny}
\and The Graduate School and University Center, City University of New York, USA\label{ny2}
\and Dipartimento di Scienze Fisiche, Universit\`a di Napoli and INFN, Sezione di Napoli, Italy\label{napoli}
}

\date{}

\maketitle
%

\begin{abstract}
We present the version 2.0 of the program package \GOSAM{}  for the automated calculation of 
one-loop amplitudes.
\GOSAM{} is devised to compute one-loop QCD and/or electroweak corrections 
to multi-particle processes within and beyond the Standard Model.
The new code contains improvements in the generation and in the
reduction of the amplitudes, performs better in computing time and numerical accuracy, 
and has an extended range of applicability.
The extended version of the ``Binoth-Les-Houches-Accord" interface 
to Monte Carlo programs  is also implemented. 
We give a detailed description of installation and usage of the code, 
and illustrate the new features in dedicated examples.
\end{abstract}

\newpage
\tableofcontents

\section{Introduction}
\label{sec:intro}
After the great achievement of discovering a new boson at the
LHC~\cite{Aad:2012tfa,Chatrchyan:2012ufa}, the primary goal is now to
study its properties in detail, and to detect the slightest hints for
possible extensions of the Standard Model.  Certainly, precise theory
predictions are indispensable to achieve this aim, which calls for
calculations at next-to-leading order (NLO) accuracy and beyond.

NLO predictions nowadays should be considered as the standard for
experimental data analysis. Ideally, matching NLO results to a parton
shower and merging different jet multiplicities should be aimed for.
However, this also requires fast and highly automated NLO tools to be
available, to be compared to a vast amount of measurements, most of
them dealing with multi-particle final states.

The development of automated NLO tools has seen tremendous progress in
the past years, leading to public codes~\cite{Hahn:2010zi,Bevilacqua:2011xh,Hirschi:2011pa,Cullen:2011ac,Badger:2012pg} 
which are able to produce multi-particle NLO predictions for user-defined processes,
or to dedicated 
frameworks~\cite{Berger:2008sj,Bredenstein:2010rs,Alioli:2010xd,Campbell:2011bn,Arnold:2011wj,Becker:2011vg,Cascioli:2011va,Actis:2012qn},
which allowed to produce an impressive collection of NLO processes.

The integrand-reduction
method \cite{Ossola:2006us,Ossola:2007bb,Ossola:2007ax} has changed
our way of addressing the decomposition of amplitudes in terms of master
integrals, whose coefficients can be determined by applying algebraic
projections to polynomial functions.

The principle of an integrand-reduction method, which is valid at any
order in perturbation
theory \cite{Mastrolia:2011pr,Badger:2012dp,Zhang:2012ce,Mastrolia:2012an,Mastrolia:2013kca},
is the underlying multi-particle pole expansion for the integrand of
any scattering amplitude, or, equivalently, a representation where the
numerator of each Feynman integral is expressed as a combination of
products of the corresponding denominators, with polynomial
coefficients.  These coefficients correspond to the residue of the integrand at the
multiple-cut. Each residue is a multivariate polynomial in the {\it
irreducible scalar products} formed by the loop momenta and either
external momenta or polarization vectors constructed out of them.

\GOSAM{} is a code which was designed to maximally exploit both the integrand
reduction for dimensionally regulated one-loop
amplitudes \cite{Ossola:2006us,Giele:2008ve}, as implemented
in \SAMURAI{} \cite{Mastrolia:2010nb}, as well as 
improved tensor reduction methods as developed in \cite{Binoth:2005ff,Heinrich:2010ax}.
The algebraic expression of
the integrands are automatically generated by means of the {\sc Golem}
technology \cite{Binoth:1999sp,Binoth:2005ff,Reiter:2009kb,Cullen:2011kv}.

The polynomial structure of the multi-particle residues is a {\it
qualitative} information that turns into a {\it quantitative}
algorithm for decomposing arbitrary amplitudes in terms of master
integrals at the integrand level. In fact, in the context of an
integrand-reduction, any explicit integration procedure is replaced by
a simpler operation like {\it polynomial fitting}, which in \SAMURAI{}
is implemented via Discrete Fourier Transform
\cite{Mastrolia:2008jb,Mastrolia:2012du,vanDeurzen:2013pja}.

\GOSAM{} produces analytic expressions for the
integrands. Because of this feature, it is suitable to be interfaced
with a new library, called \NINJA{}
\cite{Peraro:2014cba,vanDeurzen:2013saa}, 
implementing an ameliorated integrand-reduction method, where the
decomposition in terms of master integrals is achieved by Laurent
expansion through semi-analytic {\it polynomial
divisions} \cite{Mastrolia:2012bu}.  With the new reduction
algorithm, \GOSAM{}-2.0 can produce results for NLO virtual corrections
that are more accurate and less time consuming than the ones provided by version 1.0.

In this paper we present the new version 2.0 of the
program \GOSAM~\cite{Cullen:2011ac}, which has been used already to
produce a multitude of NLO predictions both
within~\cite{Greiner:2011mp,Greiner:2012im,vanDeurzen:2013rv,Gehrmann:2013aga,Luisoni:2013cuh,Hoeche:2013mua,Cullen:2013saa,vanDeurzen:2013xla,Gehrmann:2013bga,Dolan:2013rja,Heinrich:2013qaa,vanDeurzen:2013saa}
and beyond~\cite{Cullen:2012eh,Greiner:2013gca} the Standard Model.
The new version contains important improvements in speed, numerical
robustness,  range of applicability and user-friendliness.
\GOSAM{} can be linked to different Monte Carlo programs via the Binoth-Les-Houches-Accord BLHA~\cite{Binoth:2010xt},
where the extended version BLHA2~\cite{Alioli:2013nda} is also implemented. The program can be downloaded from~\cite{gosamhome}
\bcen
    \url{http://gosam.hepforge.org}
\ecen

The structure of the paper is the following.
In Section~\ref{sec:overview} we give a brief overview of the program structure. 
The new features of the program are presented in Section~\ref{sec:newfeatures}. 
Section~\ref{sec:instantuse} describes the installation and usage of \GOSAM{}, 
while in Section~\ref{sec:examples} we give examples illustrating 
some of the new features, before we conclude in Section~\ref{sec:conclusion}.
The appendices contain a commented example of an input card for convenience of the user, 
and some details about higher rank integrals.


\section{Overview of the program}
\label{sec:overview}
\begin{center}
\begin{figure}[htb]
\begin{picture}(100,85)
\put(20,-110){\includegraphics[width=15cm]{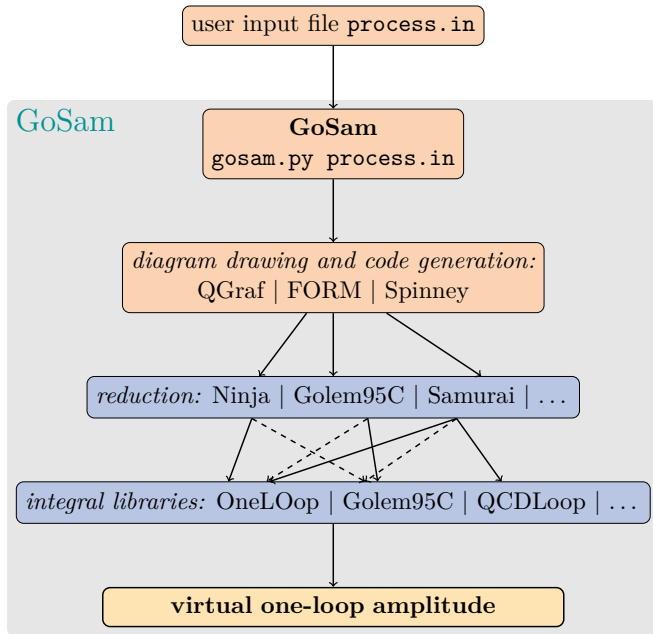}}
\end{picture}
\caption{Basic workflow of \GOSAM{}.}
\label{fig:flowchart}
\end{figure}
\end{center}

\begin{figure}[htb]
\centering
\begin{picture}(100,110)
\put(20,-30){\includegraphics[width=11cm]{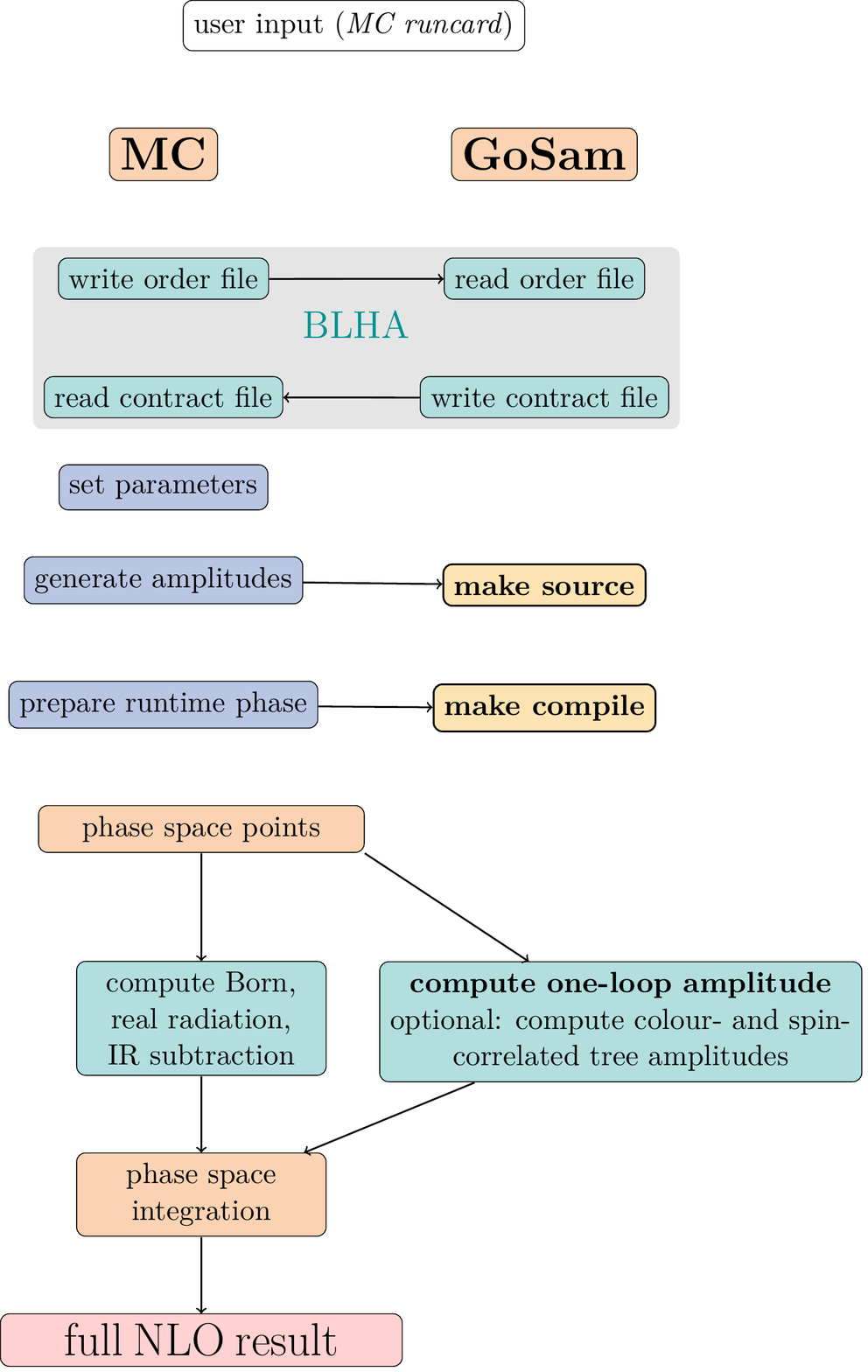}}
\end{picture}
\caption{Schematic setup for \GOSAM{} as an OLP in combination with a Monte Carlo program.}
\label{fig:olp}
\end{figure}

\GOSAM{} can be used either as a standalone code producing one-loop 
(and tree level) amplitudes, or it can be used as a {\it One Loop
Provider} (OLP) in combination with a {\it Monte Carlo} (MC) program, where the
interface is automated, based on the standards defined
in \cite{Binoth:2010xt,Alioli:2013nda}.  The main workflow of \GOSAM{}
is shown in Fig.~\ref{fig:flowchart} for  the standalone version
and in Fig.~\ref{fig:olp} for \GOSAM{} as an OLP within a Monte Carlo setup.

In the standalone version, the user will fill out a process {\it run
card} which we call {\tt process.in}, where the process is defined,
together with some options.  Then the code for the virtual amplitudes
is generated by invoking {\tt gosam.py process.in}\,.\\ After running
the above command with an appropriate run card, all the files which
are relevant for code generation will be created.  The command {\tt
make source} will invoke \QGRAF~\cite{Nogueira:1991ex}
and \FORM~\cite{Vermaseren:2000nd,Kuipers:2012rf} to generate the
diagrams and algebaric expressions for the amplitudes, using also 
\SPINNEY~\cite{Cullen:2010jv} for the spinor algebra within \FORM{}
and \HAGGIES~\cite{Reiter:2009ts} for code generation.  In
version 2.0 of \GOSAM{}, the production of optimized code however is
largely relying on the new features of \FORM~version $\geq 4$.  The
command {\tt make compile} will finally compile the produced {\tt
Fortran90} code.

In the OLP version, the information for the code generation is taken
from the order file generated by the Monte Carlo 
program. Depending on the MC, the whole generation can be invoked
automatically and steered by its setup. This is shown schematically
in Fig.~\ref{fig:olp} and explained in more detail in
Section \ref{sec:blha}.

The amplitudes are evaluated using $D$-dimensional reduction at
integrand level~\cite{Ossola:2006us,Ellis:2008ir,Mastrolia:2008jb},
which is available through two different reduction procedures and
libraries:
\SAMURAI~\cite{Mastrolia:2010nb,vanDeurzen:2013pja} and \NINJA~\cite{Mastrolia:2012bu,vanDeurzen:2013saa}. 
Alternatively, tensorial reconstruction~\cite{Heinrich:2010ax} is also
available, based on the
library \GOLEMVC~\cite{Binoth:2008uq,Cullen:2011kv,Guillet:2013msa}.
The scalar loop integrals can be evaluated using {\sc
OneLOop}~\cite{vanHameren:2010cp}, \GOLEMVC, or {\sc
QCDLoop}~\cite{vanOldenborgh:1990yc,Ellis:2007qk}.

It should be emphasized that all the  reduction- and integral libraries 
used in \gosamv{} are included
in the program package, and the installation script described in
Section~\ref{sec:install} will take care of compilation and linking,
such that the user does not have to worry about installing them
separately.
Interfacing other tensor integral libraries, such as {\tt
LoopTools}~\cite{Hahn:2010zi,Hahn:1998yk}, 
{\tt PJFRY}~\cite{Fleischer:2010sq,Fleischer:2012et} or 
{\tt Collier}~\cite{Actis:2013dfa}, 
should be straightforward, due to the modular structure of our setup.

More details about the
reduction procedures implemented in \GOSAM{} will be given in Section~\ref{sec:newfeatures}.

\section{New features}
\label{sec:newfeatures}
The version 2.0 of \GOSAM{} comes with several new features, which
lead to an improvement in speed for both the generation and the
evaluation of the amplitudes, more compact code, and more stable
numerical evaluation. Further, the range of applicability of the code
is extended, in particular to deal with effective theories and
physics Beyond the Standard Model.

\noindent We will describe some of the new features in more detail below.

\subsection{Improvements in code generation}

\subsubsection{Producing optimised code  with {\tt FORM} version 4}

While in version 1.0 of \GOSAM{} the \FORTRAN{} code for the
amplitudes was written using \HAGGIES~\cite{Reiter:2009ts}, we now
largely use the features provided by \FORM{} version
4.x~\cite{Kuipers:2012rf} to produce optimized code. This leads to more
compact code and a speed-up in amplitude evaluation of about a factor
of ten.  The option to use \HAGGIES{} is still available by setting
the extension {\tt noformopt}.

\subsubsection{Grouping/summing of diagrams which share common subdiagrams}
\label{sec:grouping_summing}
Already in the first release of \GOSAM{}, the diagrams were analyzed
according to their kinematic matrix $S_{ij}$ and grouped together
before reduction. This lead to an important gain in efficiency, both
with reduction based on integrand reduction methods, as well as with
classical tensor reduction techniques. Details about the way diagrams
are grouped can be found in~\cite{Cullen:2011ac}. This feature is
still present when \SAMURAI{} or \GOLEMVC{} are used for computing the
amplitudes.

In the new release an option called {\tt diagsum} combines diagrams
which differ only by a subdiagram into one ``meta-diagram'' to be
processed as an entity. This allows one to further reduce the number of
calls to the reduction program and therefore to increase the
computational speed. 

\begin{figure}[htb]
\centering
\includegraphics[width=0.8\textwidth]{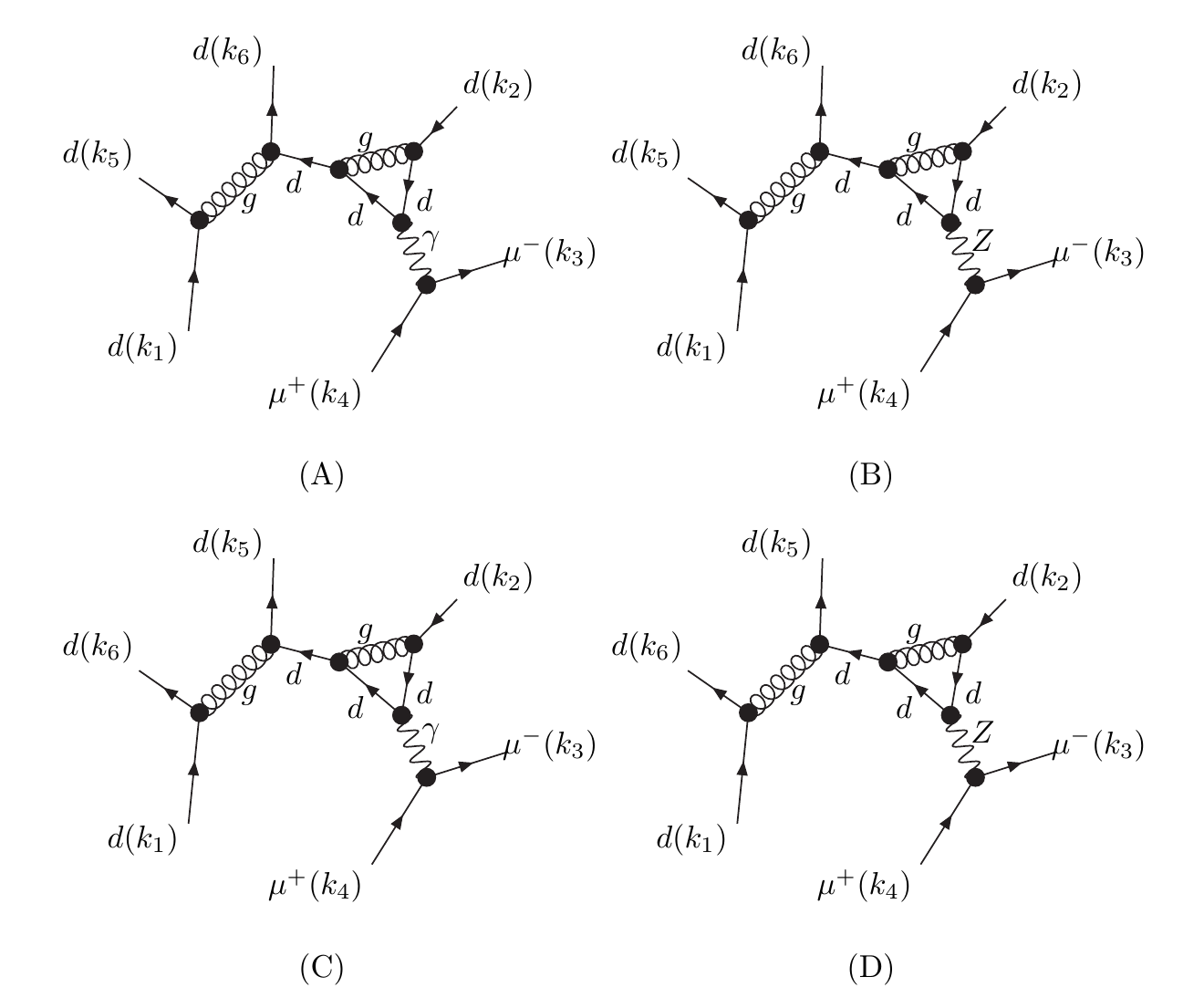}
\caption{Example of diagrams sharing a common tree part, which are 
summed when the {\tt diagsum} option is set to {\tt diagsum=true}.}
\label{fig:diagsum_tree}
\end{figure} 

\begin{figure}[htb]
\centering
\includegraphics[width=1.0\textwidth]{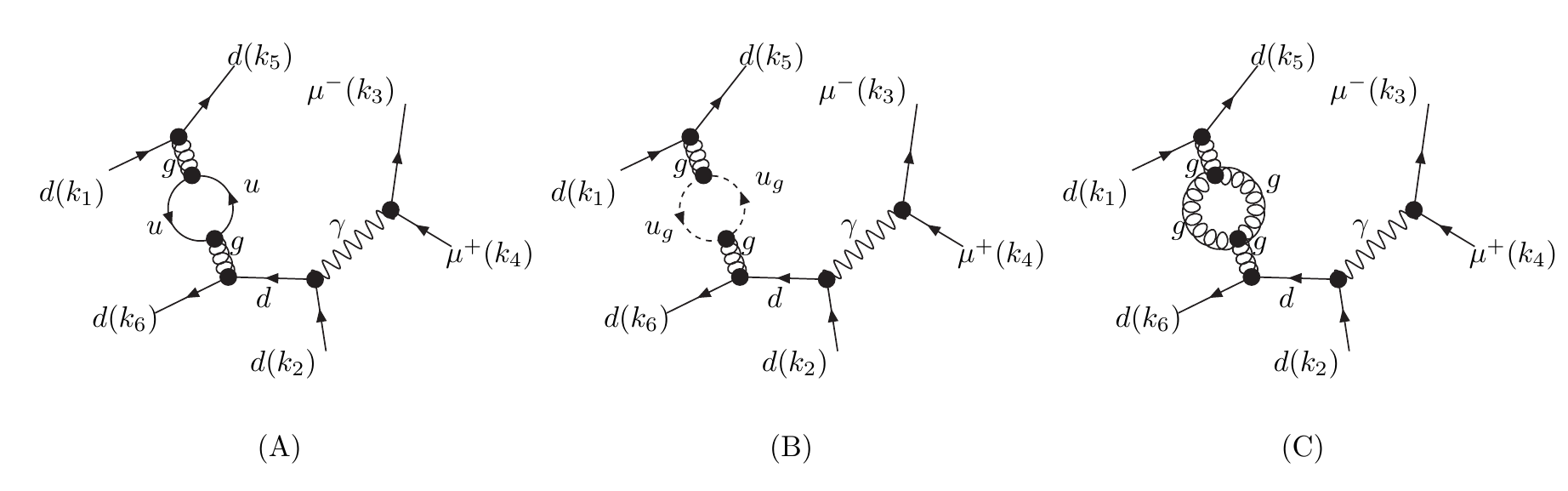}
\caption{Example of diagrams sharing a common loop propagator, 
but with different particle content in the loop, which are summed when
the {\tt diagsum} option is set to {\tt diagsum=true}.}
\label{fig:diagsum_particle}
\end{figure}

When the option {\tt diagsum} is active, diagrams which differ only by
a propagator external to the loop, as is the case e.g. for the
$Z/\gamma^\star$ propagator in QCD corrections to the production of
$Z+$jets, are summed together before being processed
by \FORM{}. Similarly, diagrams which differ only by an external tree
part, but which share exactly the same set of loop propagators, are
summed together prior the algebraic manipulation. An example is shown
Figure~\ref{fig:diagsum_tree}. Finally, diagrams which share the same
set of propagators, but have different particles circulating in the
loop, as shown in Figure~\ref{fig:diagsum_particle}, are also summed
into one ``meta-diagram''. The default setting for this option is {\tt
diagsum=true}.

\subsubsection{Numerical polarisation vectors}
\label{sec:numpolvec}
The use of numerical polarisation vectors for massless gauge bosons
(gluons, photons) is activated by default.  This means that the
various helicity configurations for the massless bosons will be
evaluated numerically, based on a unique code containing generic polarisation vectors, 
rather than producing separate code for each helicity configuration.  
To switch off this default setting,
for example if the user would like to 
optimize the choice of reference vectors for each helicity configuration,
the option {\tt polvec=explicit} should be given in the input file 
{\tt process.in}.

\subsection{Improvements in the reduction}
\label{sec:ninja}

The algebraic generation of the integrands in \GOSAM{} is tailored to the maximal exploitation of
the $D$-dimensional integrand reduction algorithm.

In the previous version, \GOSAM{}-1.0, \SAMURAI{} has been the default library for the
amplitude decomposition in terms of master integrals.
Within the original integrand reduction algorithm,
implemented in \SAMURAI{}, the determination of the unknown
coefficients multiplying the master integrals requires: 
{\it i)} to {\it sample} the numerator on a finite subset of the on-shell
solutions; 
{\it ii)} to {\it subtract} from the integrand the non-vanishing
contributions coming from  higher-point residues; 
and {\it iii)} to {\it solve} the resulting linear system of
equations. Gauss substitutions and the integrand subtractions enforce
a {\it triangular} system solving strategy, which proceeds top-down,
from the pentuple-cut to the single-cut. In this fashion, because of
the integrand subtractions, the
integrand which has to be evaluated numerically gets updated at any
level, cut-by-cut, by the subtraction of the polynomial residues
determined at the previous step.
The sampling and the determination of the coefficients in \SAMURAI{}
proceeds with a projection technique based on the Discrete Fourier
Transform \cite{Mastrolia:2008jb,vanDeurzen:2013pja}. 

In the new version \GOSAM{}-2.0, the amplitude decomposition is
obtained by a new integrand-reduction method~\cite{Mastrolia:2012bu}, 
implemented in the C++ code
\NINJA~\cite{Peraro:2014cba,vanDeurzen:2013saa}, which is the default
reduction library.

In Ref.~\cite{Mastrolia:2012bu} an improved version of the
integrand reduction method for one-loop amplitudes was presented. This
method allows, whenever the analytic dependence of the integrand on
the loop momentum is known, to extract the unknown coefficients of the
residues by performing a Laurent expansion of the integrand with
respect to one of the free loop components which are not constrained
by the corresponding on-shell conditions.

Within the Laurent expansion approach, the system of equations for the coefficients becomes
{\it diagonal}. In fact, in the asymptotic limit,
both the integrand and the higher-point subtractions exhibit the same
polynomial behaviour as the residue. Therefore one can identify the
unknown coefficients with the ones of the expansion of the integrand,
corrected by the contributions coming from higher-point residues. 
In other words, the subtractions of higher-point contributions are
replaced by corrections at the coefficient level. 
Because of the universal structure of the residues, the parametric
form of these corrections can be computed once and for all, in terms
of a subset of the higher-point coefficients.

This novel $D$-dimensional unitarity-based algorithm is lighter than
the original integrand reduction method, because less coefficients
need to be determined, and turns out to be faster and numerically
more accurate.

The integrand reduction via Laurent expansion has been implemented in
the library \NINJA{}, where the Laurent expansions of the integrands
are performed by a {\it polynomial division} between some parametric
expansions of the numerator and the uncut denominators. 
The expansions of the numerator, required by \NINJA{} as input, are
efficiently generated by \GOSAM{} using \FORM{},
after collecting the terms that do not depend on the loop momentum
into global abbreviations.

\NINJA{} and the new version of \SAMURAI{}, as well as \GOLEMVC, all distributed with the
\GOSAM{}-2.0 package, can deal with processes where the masses of the
internal particles are complex, and where the rank $r$ of the
numerator of the integrands can exceed the number $N$ of denominators
by one unit, {\it i.e.} $r \le N+1$, as it may happen e.g. in
effective theories (see also Section \ref{sec:higherrank}).

\subsubsection{The extension {\tt derive}}

The {\tt derive} feature generates code to access the tensor coefficients
of each diagram or group of diagrams individually.
While it has been among the possible keywords for the 
{\tt extensions} options in \GOSAM-1.0 already, it now has been promoted to 
be used by default in the context of  tensorial reconstruction~\cite{Heinrich:2010ax}.
It improves both the speed and the
precision of tensorial reconstruction and makes connection to other reduction methods.

The idea behind it is to compute the numerator $\mathcal{N}(q)$ 
from a Taylor series
\begin{equation}
\mathcal{N}(q)=\mathcal{N}(0)
+  q^\mu
  \frac{\partial}{\partial q_\mu}\mathcal{N}(q)\vert_{q=0}
+ \frac1{2!} q^\mu q^\nu
  \frac{\partial}{\partial q_\mu}
  \frac{\partial}{\partial q_\nu}
  \mathcal{N}(q)\vert_{q=0} + \ldots
\end{equation}
In this form one can read off a one-to-one correspondence between derivatives at
$q=0$ and the coefficients of the tensor integrals.

At a technical level, the files \texttt{helicity*/d*h*l1d.f90}
contain the routines\\
\texttt{derivative($\mu^2$, [$i_1$], [$i_2$],\dots)} and
\texttt{reconstruct\_d*(coeffs)}, where the latter is only generated in
conjunction with the extension \texttt{golem95}, and \texttt{coeffs} is
a type which comprises all coefficients of a diagram of a certain rank.
The number of optional indices $i_1$, $i_2$, \dots 
determine which derivative should be returned. The subroutine
\texttt{reconstruct\_d*} also takes into account the proper symmetrisation.


\label{sec:derive}

\subsection{Electroweak scheme choice}
\label{sec:ewchoose}
Regularisation and renormalisation within the Standard Model can be performed 
using various schemes, which also may differ in the set of  electroweak parameters 
considered as input parameters, while other electroweak parameters are derived ones.
Within \GOSAM{}-2.0, different schemes can be chosen in several
different ways by setting appropriately the flag {\tt model.options},
depending on whether the scheme might be changed after the generation
of the code or not.

By default, when the flag is not set in the input card, \GOSAM{}
generates a code which uses $\mrm{m_W}$, $\mrm{m_Z}$ and
$\mrm{\alpha}$ as input parameters, allowing however to change this in
the generated code, by setting the variable {\tt ewchoice} in the
configuration file {\tt common/config.f90} to the desired value. The user can
choose among 8 different possibilities, which are listed in
Table~\ref{tab:ewchoose}.  When the electric charge $\mrm{e}$ is set
algebraically to one, the symbol for $\mrm{e}$ will not be present in the generated amplitudes. 
This can be useful e.g. if the Monte Carlo generator used in combination with \GOSAM{}
will multiply the full result by the coupling constants at a later stage of the calculation.
In scheme number 8, $m_W$ is derived from 
a formula where both $\mrm{e}$ and $G_F$ enter. 
Thus, using $e=1$ in combination with the standard value for $G_F$
would lead to the wrong gauge boson masses.
Therefore the scheme 8 cannot be used if $\mrm{e}$ is set
algebraically to one.

\begin{table*}
\begin{center}
\small
\begin{tabular}{|c|l|l|}
\hline
ewchoice & input parameters                        & derived parameters                  \\
\hline
1        & $\mrm{G_F}$, $\mrm{m_W}$, $\mrm{m_Z}$    & $\mrm{e}$, $\mrm{sw}$              \\
2        & $\mrm{\alpha}$, $\mrm{m_W}$, $\mrm{m_Z}$ & $\mrm{e}$, $\mrm{sw}$              \\
3        & $\mrm{\alpha}$, $\mrm{sw}$, $\mrm{m_Z}$  & $\mrm{e}$, $\mrm{m_W}$             \\
4        & $\mrm{\alpha}$, $\mrm{sw}$, $\mrm{G_F}$  & $\mrm{e}$, $\mrm{m_W}$             \\
5        & $\mrm{\alpha}$, $\mrm{G_F}$, $\mrm{m_Z}$ & $\mrm{e}$, $\mrm{m_W}$, $\mrm{sw}$ \\
6        & $\mrm{e}$, $\mrm{m_W}$, $\mrm{m_Z}$      & $\mrm{sw}$                         \\
7        & $\mrm{e}$, $\mrm{sw}$, $\mrm{m_Z}$       & $\mrm{m_W}$                        \\
8        & $\mrm{e}$, $\mrm{sw}$, $\mrm{G_F}$       & $\mrm{m_W}$, $\mrm{m_Z}$           \\
\hline
\end{tabular}
\end{center}
\caption{Possible choices to select the electroweak scheme.
To simplify the notation we write the sine of the weak mixing angle as
$\mrm{sw}$. The lists of derived parameters contain only the
parameters which are computed from the input parameters and used in the expressions for the
amplitudes.}\label{tab:ewchoose}
\end{table*}

The flag {\tt model.options} in the input card allows one also to directly
set the values of the different parameters appearing in the model. If
the values of exactly three electroweak parameters are
specified, \GOSAM{} automatically takes them as input parameters. In
that case, in order to be able to switch among different schemes after
code generation, the variable {\tt ewchoose} also must be added to the
{\tt model.options} flag.

\subsection{Stability tests and rescue system}
\label{sec:rescue}

\newcommand{\gosam}{{\textsc{Go\-Sam}}}
\def\N{\mathcal{N}}
\def\S{\mathcal{S}}
\def\A{\mathcal{A}}
\newcommand{\beq}{\begin{equation}}
\newcommand{\eeq}{\end{equation}}
\newcommand{\ninja}{{\textsc{Ninja}}}
\newcommand{\bite}{\begin{itemize}}
\newcommand{\eite}{\end{itemize}}

Within the context of numerical and semi-numerical techniques, we should be able to assess in real time, for each phase space point, the level of precision of the corresponding one-loop matrix element. Whenever a phase space point is found in which the quality of the result falls below a certain threshold, either the point is discarded or the evaluation of the amplitude is repeated by means of a safer, albeit less efficient procedure. This procedure is traditionally called ``rescue system''.

Apart from improvements in the stability of the reduction itself, which are provided by the new versions of \SAMURAI{} and \GOLEMVC, and in particular by the new reduction algorithm \NINJA,  the new version of \GOSAM{} also has a more refined rescue system as compared to version 1.0. 

Looking at the literature, we observe that various techniques for detecting points with low precision have been implemented within the different automated tools for the evaluation of one-loop virtual corrections.

A first commonly used approach relies on the comparison between the numerical values of the infrared pole coefficients computed by the automated tool with their known analytic results dictated by the universal behaviour of the infrared singularities~\cite{Catani:2000ef}. We refer to this as the {\it pole test}. 

The main advantages of this method are its broad applicability to all amplitudes and the fact that it requires a negligible additional computation time. However, since not all integrals which appear in the reconstruction of the amplitude give a contribution to the double and single poles, this method often provides an overestimate of the precision, which might result in keeping  phase space points whose finite part is less precise than what is predicted by the poles.

Different techniques have been proposed that target directly the precision of the finite part. Using the symmetry properties of scattering amplitudes under scaling of all physical scales, or alternatively the invariance under rotation of the momenta, we can build pairs of points that should provide identical results, both for the finite parts and for the poles, and use the difference between them as an estimator of the precision. 

The {\it scaling test}~\cite{Badger:2010nx}, is based on the properties of scaling of scattering amplitudes when all physical scales (momenta, renormalisation scale, masses) are rescaled by a common multiplicative factor $x$. As shown in~\cite{Badger:2010nx}, this method provides a very good correlation between the estimated precision and the actual precision of the finite parts.

The {\it rotation test}~\cite{vanDeurzen:2013saa} exploits the invariance of the scattering amplitudes under an azimuthal rotation about the beam axis, namely the direction of the initial colliding particles. Whenever the initial particles are not directed along the beam axis, one can perform a rotation of all particles by an arbitrary angle in the space of momenta. A validation of this technique, and the corresponding correlation plots, has been presented in~\cite{vanDeurzen:2013saa}.

While  the {\it scaling} and the {\it rotation test} provide a more reliable estimate of the precision of the finite parts that enter in the phase space integration, their downside is that they require two evaluations of the same matrix element, therefore leading to a doubling in the computational time.

Additional methods have been proposed, within the context of integrand-reduction approaches, which target the relations between the coefficients before integration, namely the reconstructed algebraic expressions for the numerator function before integration, known as  $\N=\N$ tests~\cite{Ossola:2007ax, Mastrolia:2010nb}.
This kind of tests can be applied to the full amplitude (global   $\N=\N$ test) or individually within each residue of individual cuts (local $\N=\N$ test). The drawback of this technique comes from the fact that the test is applied at the level of individual diagrams, rather than on the final result summed over all diagrams, making the construction of a rescue system quite cumbersome. 

For the precision analysis contained in \gosamv, and to set the trigger for the rescue system, we decided to employ  a hybrid method, that takes advantage of the computational speed of the {\it pole test}, combined with the higher reliability of the {\it rotation test}.  This hybrid method requires setting three different thresholds.
After computing the matrix elements, \gosamv{} checks the precision  $ \delta_{pole}$ of the single pole with the {\it pole test}. Comparing the single pole 
$\S_{IR}$ that can be obtained from the general structure of infrared singularities and the one provided by  \gosamv, which we label $\S$, we define  
\beq \label{eq:exd}
\delta_{pole} = \left | \frac{ \S_{IR} - \S{} }{ \S_{IR}} \right |\, .
\eeq
The corresponding estimate of the number of correct digits in the result is provided by  $P_{pole}= - \log_{10} (\delta_{pole})$. This step does not require any increase in computational time. The value of $ P_{pole}$ is then compared with two thresholds $ P_{high}$ and $ P_{low}$. 

If $P_{pole} >  P_{high}$ the point is automatically accepted. Given the high quality of the computed pole, the finite part is very unlikely to be so poor that the point should be discarded.

If $P_{pole} <  P_{low}$ the point is automatically discarded, or sent to the rescue system. If already the pole has a low precision, we can expect the finite part to be of the same level or worse.
 
In the intermediate region where $ P_{high} > P_{pole} >  P_{low}$, it is more difficult to determine the quality of the result solely based on the 
pole coefficients. 
Only in this case the point is recalculated using the {\it rotation test}, which requires additional computational time. 

If we call the finite part of the amplitudes evaluated before and after the rotation $\A^{\rm fin}$ and  $\A_{rot}^{\rm fin}$ respectively,  we can define the error $ \delta_{rot}$ estimated with the rotation  as  
\beq  \label{eq:errd} \delta_{rot} =  2 \left  |\frac{ A_{rot}^{\rm fin} - A^{\rm fin} }{ A_{rot}^{\rm fin} + A^{\rm fin}} \right  |\, . \eeq
and the corresponding estimate on the number of correct digits as $P_{rot} = - \log_{10} (\delta_{rot})$.
$P_{rot}$ provides a reliable estimate of the precision of the finite part~\cite{vanDeurzen:2013saa}, and can be compared with a threshold $P_{set}$ to decide whether the point should be accepted or discarded. 

The values of the three thresholds $ P_{high} $,  $P_{low}$ and $P_{set}$ can be chosen by the user, to adjust the selection mechanism to the fluctuations in precision which occur between different processes. In the input card, $ P_{high} $,  $P_{low}$ and $P_{set}$ correspond to 
\texttt{PSP\_chk\_th1}, \texttt{PSP\_chk\_th2} and \texttt{PSP\_chk\_th3}, 
respectively, see appendix \ref{sec:appendix}.
It is worth to notice that the {\it rotation test} can be bypassed simply by setting the initial thresholds $P_{high}= P_{low}$. In this case the selection is performed solely on the basis of the {\it pole test}.

\subsection{New range of applicability}

\subsubsection{Higher rank integrals}
\label{sec:higherrank}

The libraries \NINJA, \GOLEMVC{} and \SAMURAI{} all support integrals
with tensor ranks $r$ exceeding the number of propagators $N$.  Such
integrals occur for example in effective theories (a prominent example
is the effective coupling of gluons to the Higgs boson), or in
calculations involving spin-two particles beyond the leading order.
These extensions are described in detail in
Refs.~\cite{vanDeurzen:2013pja,Guillet:2013msa,Mastrolia:2012bu} and
are contained in the distribution of \GOSAM-2.0.  The additional
integrals will be called automatically by \GOSAM{} if they occur in an
amplitude, such that the user can calculate amplitudes involving
higher rank integrals without additional effort.  \NINJA{}
and \SAMURAI{} provide higher rank integrals for rank
$r=N+1$, version 1.3 of \GOLEMVC{}~\cite{Guillet:2013msa} 
provides higher rank integrals and the tensorial reconstruction routines 
up to $r=N+1$ for $N\leq 6$, 
as well as form factors  up to rank ten for $N\leq 4$.
More details about the higher rank
integrals are given in Appendix \ref{sec:appendixB}.

\subsubsection{Production of colour-/spin correlated trees}

\GOSAM{} can also generate  tree level amplitudes in a spin- and colour-correlated form.
Colour correlated matrix elements are defined as
\begin{equation}
 C_{ij}=\bra{{\cal M}}\textbf{T}_{i}\textbf{T}_j \ket{{\cal M}}\;,
\end{equation}
and we define spin-correlated matrix elements  as
\begin{equation}
 S_{ij}=\bra{{\cal M},-}\textbf{T}_{i}\textbf{T}_j \ket{{\cal M},+}\;.
\end{equation}
The spin-correlated matrix element (as well as the colour correlated
matrix element) contains implicitly the sum over all non-specified helicities, while
only the helicities with the indices $i$ and $j$ are fixed,
i.e.  
\begin{eqnarray} &&\langle {\cal M}_{i,-} |{\mathbf T}_i\cdot
{\mathbf T}_j |{\cal M}_{i,+}\rangle =\\
&&\sum_{\lambda_1,...,\lambda_{i-1},\lambda_{i+1},...,\lambda_n}
\langle {\cal M}_{\lambda_1,...,\lambda_{i-1},-,\lambda_{i+1},...,\lambda_n} |
{\mathbf T}_i\cdot {\mathbf T}_j | 
{\cal M}_{\lambda_1,...,\lambda_{i-1},+,\lambda_{i+1},...,\lambda_n}\rangle \;. \nonumber
\end{eqnarray}
These matrix elements are particularly useful in combination with
Monte Carlo programs which use these trees to build the dipole
subtraction terms for the infrared divergent real radiation part. With
these modified tree level matrix elements \GOSAM{} is able to generate
all necessary building blocks for a complete NLO calculation.\\ Such a
setup has been used successfully in combination with the framework of
{\sc Herwig++/Matchbox}~\cite{LesHouches2013,Bellm:2013lba,Platzer:2011bc}.

\subsubsection{Support of complex masses}
\label{sec:complexmasses}
The integral libraries contained in the \GOSAM{} package as well as the \GOSAM{} 
code itself fully support complex masses. 
The latter are needed  for the treatment of 
unstable fermions and gauge bosons  via the   introduction of the corresponding decay width. 
A fully consistent treatment of complex
$W$- and $Z$-boson masses requires the use of the complex mass scheme~\cite{Denner:2005fg}.
According to this scheme the boson masses become
\begin{equation}
 m_{V}^2 \to \mu_{V}^2 = m_{V}^2 -i m_{V} \Gamma_{V},\quad V=W,Z\;.
\end{equation}
Gauge invariance requires that the definition of the weak mixing angle has to be modified accordingly:
\begin{equation}
 \cos^2\theta_W = \frac{\mu_W^2}{\mu_Z^2}\;.
\end{equation}
 To make use of the complex mass scheme, we introduce two new model files, \texttt{sm\_complex}
 and \texttt{smdiag\_complex}, which contain the Standard Model with complex mass scheme, the first
 with a full CKM matrix, the latter with a diagonal unit matrix for the CKM matrix.
 An example dealing with a complex top quark mass is given in Section~\ref{sec:examples}.


\section{Installation and usage}
\label{sec:instantuse}

\subsection{Installation}
\label{sec:install}
The user can download the code either as a tar-ball
    or from a subversion repository at
\bcen
    \url{http://gosam.hepforge.org}
\ecen

\noindent The installation of \gosamv{} is very simple when using the installation script.
The latter can be downloaded by
\bfl
 {\tt wget } \url{http://gosam.hepforge.org/gosam-installer/gosam\_installer.py}
\efl
By default \GOSAM{} will be installed into a subfolder {\tt ./local}
of the directory where the installation script is saved. A different
path can be specified using the 
\bfl
{\tt --prefix=PATH\_where\_to\_install}
\efl
option. To run the script the user should execute the following commands
\bfl
{\tt chmod +x gosam\_installer.py}\\
{\tt ./gosam\_installer.py  [--prefix=...]}
\efl
 or
\bfl
{\tt python gosam\_installer.py   [--prefix=...]}
\efl
Upon installation, the installer will ask some questions, which are
described in more detail in the manual~\cite{gosamhome}, which also can be downloaded from the  webpage
given above. 

To use the default installation all the questions can be ``answered" by pressing the {\tt ENTER} key.

In particular, the installer will check if  \QGRAF~\cite{Nogueira:1991ex} 
and {\tt FORM}~\cite{Vermaseren:2000nd,Kuipers:2012rf}
already exist on the system.
If they are not found, one can either 
press {\tt ENTER} to have them installed by the script, or provide a path to
the binary (tab-completion can be used).
If they are found, their version is checked, 
and if needed the installation
of a version which has been tested to run with \GOSAM{} is suggested.

As soon as all questions are answered, the main installation process
will start.  The components will be downloaded, built and
installed. The whole procedure can take about 10-30 minutes.

At the end, a script {\tt gosam\_setup\_env.sh} will be created in the
{\tt bin/} subdirectory of the install location, which will set
(temporarily) all environment variables as soon as the script is
sourced into a shell (with the command {\tt source
[path]/gosam\_setup\_env.sh}).  The installer also gives a
recommendation how these environment variables can be set permanently.
The script can be used in all {\tt tcsh}- and {\tt bash}-compatible
shells.

All files which have been installed are tracked in the {\tt
installer-log.ini} file. It is important to keep this file and the
install script. They are needed to update and uninstall \GOSAM. For
the default installation, internet access is required.

\medskip

The program \GOSAM{} is designed to run in any modern Unix-like environment (Linux, Mac).
The system requirements are
\python~($\geq2.6$), 
a \FORTRAN\, compiler (gfortran or ifort), 
a C/C++ compiler (gcc/icc), and (GNU) {\tt make}. 
Compatibility with gcc versions 4.2.--4.9 as well as  {\tt clang} has been tested.
By default, \GOSAM{} uses the gfortran/gcc  compilers from the GNU Compiler Suite.
To use an Intel compiler  (ifort/icc), the {\tt --intel} option can be used.
Specific paths to the compilers can be provided using the {\tt --fc, --cc, --cxx}  options.

\noindent All further options can be listed by invoking the installation script with the flag {\tt help}:
\bfl
{\tt gosam\_installer.py --help}.
\efl

\subsection{Using \GOSAM}
\label{sec:usage}


We first start describing the use of \GOSAM{} in the standalone
version.  For the use in combination with a Monte Carlo program, based
on the BLHA interface, we refer to Section~\ref{sec:blha}.

In order to generate the matrix element for a given process the user
should create a process specific setup file, \texttt{process.in},
which we call {\em process card}.  An example process card for the
process $e^+e^-\to t\bar{t}$ is given in Appendix~\ref{sec:appendix},
where we explain each entry in detail.

It is recommended to generate and modify a template file for the
process card. This can be done by invoking the shell command
\bcen
{\tt gosam.py --template process.in}
\ecen
which generates the file \texttt{process.in} with some documentation for
all defined options. The options are filled with the default values,
and some paths are set by the installation script. User-defined
options changing the default values can also be set in a global
configuration file. The script will search in the \gosamv{}
directory, in the user's home directory and in the current working
directory for a file named `\texttt{.gosam}' or `\texttt{gosam.in}'.

In order to generate the Fortran code for the process specified in the
input card one needs to invoke
\bcen
{\tt
 gosam.py process.in
}
\ecen

\subsection*{\bf Structure of the generated code}

The generated process directory will have the following sub-directory
structure:
\begin{itemize}
\item{\tt codegen}: This directory contains files which are only
relevant for code generation. These files will therefore not be
included in a tar-ball created with \texttt{make dist}.

\item{\tt common}: Contains {\tt Fortran} files which are common to all helicity
amplitudes and to the constructed matrix element code. 
The file {\tt config.f90} contains some global  settings, the file {\tt model.f90}
contains the definitions and settings for the model parameters.
This directory is always compiled first.

\item{\tt doc}: Contains  files which are necessary for creating
\texttt{doc/process.pdf}, which displays all Feynman diagrams of the Born level and one-loop amplitude, 
together with colour and helicity info.

\item{\tt helicity*}: These directories contain all files for a specific
helicity amplitude. The labeling of the helicities can be found in
\texttt{doc/process.pdf}. Before invoking \texttt{make source}, 
this directory only contains the makefiles. After the full code
generation, for each diagram three classes of files are created. The
basic algebraic expressions for the individual one-loop diagrams are
contained in the files {\tt d*h*l1.txt} in an optimized format. The
files {\tt d*h*l1.prc} contain the expressions of the numerators as 
polynomials in the loop momentum. The corresponding {\tt Fortran} files
are {\tt d*h*l1.f90} and {\tt abbrevd*h*.f90}, where the latter
contains the abbreviations. 

Files generated with the {\tt derive} option (see Sec.~\ref{sec:derive}) are
named {\tt d*h*l1d.*}, while the input files
for \NINJA{} (see Sec.~\ref{sec:ninja}) are named {\tt d*h*l1*.*}. For more
details we refer to the manual~\cite{gosamhome}.




\item{\tt matrix}: Contains the code to combine
the helicity amplitudes into a matrix element. Here one also finds the
test program \texttt{test.f90}. The files in this directory are always
compiled last.

\item Further, there are some files in the main process directory, for example
the Born/loop diagram files generated by \QGRAF, called {\tt
diagrams-[0/1].hh}, or the model file {\tt model.hh}.

\end{itemize}

\subsection*{{\bf Conventions}}

In the case of QCD corrections, the tree-level
matrix element squared can be written as
\begin{equation}\label{eq:amp0:def}
\vert\mathcal{M}\vert_{\text{tree}}^2=\mathcal{A}_0^\dagger\mathcal{A}_0=
(g_s)^{2b}\cdot a_0\;.
\end{equation}
The fully renormalised matrix element at one-loop level, i.e. the
interference term between tree-level and one-loop amplitudes, can be written as
\begin{multline}\label{eq:amp1:def}
\vert\mathcal{M}\vert_{\text{1-loop}}^2=
\mathcal{A}_1^\dagger\mathcal{A}_0+
\mathcal{A}_0^\dagger\mathcal{A}_1=
2\cdot\Re(\mathcal{A}_0^\dagger\mathcal{A}_1)=
\\
\vert\mathcal{M}\vert^2_{\text{bare}}
+\vert\mathcal{M}\vert^2_{\text{ct, $\delta m_Q$}}
+\vert\mathcal{M}\vert^2_{\text{ct, $\alpha_s$}}
+\vert\mathcal{M}\vert^2_{\text{wf, g}}
+\vert\mathcal{M}\vert^2_{\text{wf, Q}}=\\
\frac{\alpha_s(\mu)}{2\pi}\frac{(4\pi)^\varepsilon}{\Gamma(1-\varepsilon)}
\cdot (g_s)^{2b}\cdot\left[%
c_0+\frac{c_{-1}}{\varepsilon}+\frac{c_{-2}}{\varepsilon^2}
+\mathcal{O}(\varepsilon)%
\right]\;.
\end{multline}
In the default case the flag {\tt nlo\_prefactors} has the value zero, 
which means that a call to the subroutine \texttt{samplitude} returns an array
consisting of the four numbers $(a_0, c_0, c_{-1}, c_{-2})$, 
in this order,  
with the prefactors extracted as given above. 
In the case of electroweak corrections {\tt nlo\_prefactors=0}
has the same meaning, except that $\alpha_s$  is replaced by $\alpha$.
If the flag {\tt nlo\_prefactors} has the value one, a factor of 
$1/8\pi^2$ instead of $\alpha_s/2\pi$ (respectively $\alpha/2\pi$ in the EW case)
has been extracted from the numerical result, 
while for {\tt nlo\_prefactors=2} all the prefactors are included in the 
numerical result.

The average over initial state colours and helicities is included 
in the default setup. 
In cases where the process is loop induced, i.e. the tree level amplitude is absent, 
the program returns the values for $\mathcal{A}_1^\dagger\mathcal{A}_1$,  
where a factor of $$\left(\frac{\alpha_s(\mu)}{2\pi}\frac{(4\pi)^\varepsilon}{\Gamma(1-\varepsilon)}\right)^2$$ has been pulled out.

After UV renormalisation has been performed, only IR-singularities remain
in the virtual matrix element. The coefficients of the latter can be checked 
using the routine \texttt{ir\_sub\-trac\-tions}. This routine
constructs the pole parts of the dipole subtraction terms and 
returns a vector of length two, containing the coefficients of the
single and the double pole, which should be equal to $(c_{-1}, c_{-2})$.

\subsection{Interfacing to Monte Carlo programs}
\label{sec:blha}
The interface of \GOSAM{} with a Monte Carlo event generator program 
is based on the Binoth-Les Houches Accord (BLHA) standards.
\GOSAM{}-2.0 supports both BLHA1~\cite{Binoth:2010xt}
and BLHA2~\cite{Alioli:2013nda}.
Certainly, a dedicated interface without using the BLHA is also possible,
and such an interface with {\sc MadGraph/MadDipole/MadEvent}~\cite{Stelzer:1994ta,Frederix:2008hu,Frederix:2010cj,Alwall:2007st} 
has been built and applied successfully in various phenomenological 
applications~\cite{Cullen:2013saa,Gehrmann:2013aga,Gehrmann:2013bga,Greiner:2013gca,Cullen:2012eh}.

If \GOSAM{} is used as a One Loop Provider (OLP), 
the Monte Carlo program is steering the different stages of the calculation, 
in particular the 
phase space integration and the event generation, as illustrated in Fig.~\ref{fig:olp}.
Therefore, the user frontend  will depend on the user interface of the Monte Carlo program. 
The latter will call \GOSAM{}  at runtime  
to provide the corresponding value of the one-loop amplitude at the given phase space points. 

A number of phenomenological results 
produced by combining \GOSAM{} with various Monte Carlo programs 
can be found in the literature, e.g 
in combination with 
{\sc Sherpa}~\cite{vanDeurzen:2013rv,Hoeche:2013mua,vanDeurzen:2013xla,Cullen:2013saa,Heinrich:2013qaa}, 
{\sc PowHeg}~\cite{Luisoni:2013cuh}, 
{\sc Herwig++/Matchbox}~\cite{LesHouches2013}. 

\noindent Examples how to run \GOSAM{} with {\sc Sherpa} can also be found on the 
{\sc Sherpa} manual webpage~\cite{SherpaDocu} and on the \GOSAM{} 
process packages webpage~\cite{gosamproc}.
For the interface with {\sc PowHeg}, a detailed description can be found in the appendix of
Ref.~\cite{Luisoni:2013cuh}. 
The interface with {\sc Herwig++/Matchbox} is described in~\cite{LesHouches2013}.

\subsection{Using external model files}
\label{sec:model}
The \GOSAM{}-2.0 package comes with the built-in model files 
{\tt sm}, {\tt smdiag}, {\tt smehc}, 
{\tt sm\_complex}, {\tt smdiag\_complex}, 
where the latter two are needed in the case of complex masses and couplings, 
see Section \ref{sec:complexmasses}. 
The model files {\tt smehc}  contain the effective Higgs-gluon couplings.

Other models can be imported easily in the {\tt UFO} (Universal FeynRules Output)~\cite{Degrande:2011ua} format.
The model import in the {\tt UFO} format can be used in the standalone as well as the OLP 
mode of \GOSAM, where both the BLHA1 and BLHA2 standards are supported for the syntax of the model import.

A model description in the {\tt UFO} format consists of a \texttt{python} package
which the user can either generate using {\sc FeynRules}~\cite{Christensen:2008py,Alloul:2013bka} 
or write himself and 
store in any directory. In order to import the model into \GOSAM{} one needs
to set the \texttt{model} variable in the {\em process card} (line 5 in the example process card of Appendix \ref{sec:appendix}),  
specifying the keyword \texttt{FeynRules} in front of the path pointing to the \texttt{python} files 
defining the model. 
For example, if  we assume that
the model description is in the directory \texttt{\$HOME/models/MSSM\_UFO}, the 
{\em process card} should contain the line \\
{\tt model= FeynRules,\$HOME/models/MSSM\_UFO}\,.

The import of model files generated by {\tt LanHEP}~\cite{Semenov:2010qt} is also supported. 
More details about the import from  {\tt LanHEP} are given in the \GOSAM{}-2.0 manual~\cite{gosamhome}. 

It should be pointed out that \GOSAM{}-2.0 provides automatic renormalisation only for 
QCD corrections. If external model files are used, as well as in the case of electroweak corrections,
including the correct renormalisation is at the responsibility of the user.

The {\tt examples} directory of the \GOSAM{}-2.0 package contains several examples 
for the import of model files, both in {\tt UFO} and in {\tt LanHEP} format. 
The subdirectory {\tt examples/model} contains model files for the MSSM (as well as for the SM)
in both {\tt UFO} and {\tt LanHEP} format.
A concrete  BSM example is discussed in Subsection \ref{sec:BSMexample}.

\section{Examples}
\label{sec:examples}
\subsection{$gg\to H+$1 jet in the heavy top mass limit}
Recently \GOSAM{} was used to compute the virtual corrections for the
production of a Higgs boson in association with 2 and 3
jets~\cite{vanDeurzen:2013rv,Cullen:2013saa} in the infinite top-mass
limit. As an example for this type of processes, where a special model
file is needed, containing the Feynman rules for the effective
vertices, which furthermore give rise to higher rank loop integrals,
we consider here the process $gg\to H\,g$.

An example {\it process card} for the generation of this process, and a test routine
comparing a phase space point with results from analytical amplitude representations
is provided among the examples of the \GOSAM{}-2.0 distribution. In
the following we will refer to that example to describe some feature
of this process.

In order to compute amplitudes using the effective gluon-gluon-Higgs
vertices, the model {\tt smehc} has to be used. This model contains
also the effective vertex for the Higgs boson decaying to two
photons. When setting the powers of the strong coupling using the {\tt
  order} flag, one has to remember that the effective vertex counts as
two powers of the strong coupling. To compute the virtual corrections
for $H+1$ jet we therefore have to set {\tt order=QCD, 3, 5}.\\ \\
The inclusion of the effective gluon-gluon-Higgs coupling at NLO also requires 
corrections of the Wilson coefficient. At NLO the Wilson coefficient
is given by \cite{Chetyrkin:1997un}
\begin{equation}
 C_1= -\frac{\alpha_s}{3\pi}\left(1+\frac{\alpha_s}{\pi}\frac{11}{4}\right)\;,
\end{equation}
where the effective Lagrangian is given by
\begin{equation}
 {\cal L}_{\mathrm{eff}}=-\frac{C_1}{4v}HG^{a, \mu \nu}G^a_{\mu \nu}\;.
\end{equation}
The {\tt smehc} model file also contains the effective vertex for 
the Higgs decay into a pair of photons via top- and $W$-loops. For the 
vertex factor we use the formula given by {\sc FeynRules}~\cite{Christensen:2008py}:
\begin{eqnarray}
 g_{\gamma \gamma H}&&=\frac{47 e^2}{72 \pi^2 v}\left(1
  -\frac{14}{705}x_t^2
  -\frac{2}{987}x_t^4 
  +\frac{33}{470}x_W^2
  +\frac{57}{6580}x_W^4
  +\frac{87}{65800}x_W^6
\right.\nonumber\\
&&\left.
  +\frac{41}{180950}x_W^8
  +\frac{5}{119756}x_W^{10} 
  +\frac{213}{26346320}x_W^{12}
  \right)\;,
\end{eqnarray}
where $x_t=\frac{m_H}{m_t}$, $x_W=\frac{m_H}{m_W}$ and the
corresponding effective Lagrangian is given by
\begin{equation}
 {\cal L}_{\mathrm{eff}}=-\frac{1}{4}g_{\gamma \gamma H}H F^{\mu \nu}F_{\mu \nu}\;.
\end{equation}
Table~\ref{tab:ggHgresult} 
contains numerical results for $gg\to Hg$  at the phase space point shown 
in Table~\ref{tab:ggHgpoint}, where we have used $m_H=125$\,GeV, 
$v^2=\frac{1}{\sqrt{2}G_F}$, $G_F=1.16639\times 10^{-5}$\,GeV$^{-2}$.

\begin{table*}
\begin{center}
\begin{kinematics}
$g$ & 298.17848024073913 & 0 & 0 &  298.17848024073913 \\
$g$ & 298.17848024073913 & 0 & 0 & -298.17848024073913 \\
$H$ & 311.27885554899825 & -282.56832327194081 & -20.783785017815998 &  31.507187680134837 \\
$g$ & 285.07810493248002 &  282.56832327194081 &  20.783785017815998 & -31.507187680134837 \\
\end{kinematics}
\end{center}
\caption{Kinematic point used for $gg\to Hg$. The Higgs boson mass is set to $m_H=125$ GeV.}\label{tab:ggHgpoint}
\end{table*}

\begin{table*}
\begin{center}
\small
\begin{tabular}{|lll|}
\hline 
\multicolumn{3}{|c|}{\small Results for $gg\to Hg$ with the kinematic point from Table~\ref{tab:ggHgpoint}.}\\
\hline
 & \multicolumn{1}{l}{\small \GOSAM{} result} & \multicolumn{1}{l|}{\small MCFM result}\\
\hline
$a_{0}$      & $\phantom{-}7.274563870476018\times10^{-4}$ & $\phantom{-}7.2745638706144032\times10^{-4}$ \\
$c_0/a_0$    & $\phantom{-}13.195495732443156$             & $\phantom{-}13.195495732443119$             \\
$c_{-1}/a_0$ & $\phantom{-}12.160134391476801$             & $\phantom{-}12.160134391476900$             \\
$c_{-2}/a_0$ & $-8.9999999999999698$                       & $-9.0000000000000000$            \\
\hline
\end{tabular}
\end{center}
\caption{Result for Born and virtual amplitude 
including the QCD corrections to
 $gg\to Hg$. The renormalisation scale is set to $\mu=m_H=125$ GeV.}\label{tab:ggHgresult}
\end{table*}

\subsection{Single top production}

An example containing complex masses in the loop propagators
is the so called s-channel single top quark production, where the
top quark has a width $w_t=1.5$\,GeV. In Table~\ref{tab:stopresult} 
we give numerical results for the subprocess 
$u\bar{d}\to \nu_e e^+ b\bar{b}$ at the phase space point given
in Table~\ref{tab:stoppoint}. The $b$-quarks are taken to be massless
and the comparison has been performed against the HELAC-NLO code~\cite{Bevilacqua:2011xh}.

\begin{table*}[h!]
\begin{center}
\begin{kinematics}
$u$       & 250 & 0 & 0 &  250 \\
$\bar{d}$ & 250 & 0 & 0 & -250 \\
$\nu_e$   & 147.53211468467353 &  24.970405230567895 & -18.431576028372117  &  144.23065114968881 \\
$e^+$     & 108.70359662136400 &  103.25573902554709 & -0.54846846595840537 &  33.976807664202191 \\
$b$       & 194.06307653413651 & -79.895963003674623 &  7.4858666717648710  & -176.69486288452802 \\
$\bar{b}$ & 49.701212159825850 & -48.330181252440347 &  11.494177822565669  & -1.5125959293629665 \\
\end{kinematics}
\end{center}
\caption{Kinematic point used for $u\bar{d}\to \nu_e e^+ b\bar{b}$. The $W$-boson and top-quark mass and width are set to $m_W=80.25$~GeV, $w_W=0$, $m_t=170.9$~GeV and $w_t=1.5$~GeV.}\label{tab:stoppoint}
\end{table*}

\begin{table*}[h!]
\begin{center}
\small
\begin{tabular}{|lll|}
\hline 
\multicolumn{3}{|c|}{\small Results for $u\bar{d}\to \nu_e e^+ b\bar{b}$ with the kinematic point from Table~\ref{tab:stoppoint}.}\\
\hline
 & \multicolumn{1}{l}{\small \GOSAM{} result} & \multicolumn{1}{l|}{\small HELAC-NLO result}\\
\hline
$a_{0}$      & $\phantom{-}6.7779888808717541\times10^{-13}$ & $\phantom{-}6.7779888808718329\times10^{-13}$ \\
$c_0/a_0$    & $\phantom{-}8.8976474517729294$              & $\phantom{-}8.8976474517739739$             \\
$c_{-1}/a_0$ & $-4.9124524216371341$                        & $-4.9124524216370293$      \\
$c_{-2}/a_0$ & $-5.3333333333333393$                        & $-5.3333333333333073$      \\
\hline
\end{tabular}
\end{center}
\caption{Result for Born and virtual amplitude 
including the QCD corrections to
 $u\bar{d}\to \nu_e e^+ b\bar{b}$. The renormalisation scale is set to $\mu=m_t=170.9$ GeV.}\label{tab:stopresult}
\end{table*}

\subsection{Graviton production within models of large extra dimensions}
\label{sec:BSMexample}

As an example for the usage of \GOSAM{} with a model file different
from the Standard Model we consider the QCD corrections to graviton
production in ADD models~\cite{ArkaniHamed:1998rs,Antoniadis:1998ig} with large extra
dimensions (LED).  The corresponding model files in {\tt
  UFO}\,\cite{Degrande:2011ua} format, which we generated using {\sc
  FeynRules}\,\cite{Christensen:2008py,Alloul:2013bka}, are located in
the subdirectory {\tt examples/model/LED\_UFO}.  To import new model
files within the \GOSAM{} setup, the user should specify the path to the
model file in the {\it process card}.  In the given example, this
already has been done, i.e.  the {\it process card} contains the
line {\tt model=FeynRules,[gosampath]/examples/model/LED\_UFO}.

\noindent The example process we included in the \GOSAM{}-2.0
distribution is $u\bar{u}\to G\to \gamma\gamma$, where $G$ denotes a
graviton, and the program calculates the virtual QCD corrections.
Note that this example also involves integrals where the rank exceeds the number of propagators, 
due to the spin-2 nature of the graviton.
Running {\tt make test} in the subdirectory {\tt examples/uu\_graviton\_yy}
should produce the result shown in
Table~\ref{tab:ledresult}, using the phase space point given in Table~\ref{tab:ledpoint}.  
The full process,
including an additional jet, has been calculated
in~\cite{Greiner:2013gca}, where we refer to for details about the
parameter settings.


\begin{table*}
\begin{center}
\begin{kinematics}
$u$        & $250$ &  0   & 0 &  $\phantom{-}250$ \\
$\bar{u}$  & $250$ &  0   & 0 &  $-250$ \\
$\gamma_1$ & $250$ &  $\phantom{-}218.30931500994714$ & $ -29.589212828575324$
& $\phantom{-}118.17580743990260$ \\
$\gamma_2$ & $250$ & $-218.30931500994714$            & $
\phantom{-}29.589212828575324 $ & $-118.17580743990260$ \\
\end{kinematics}
\end{center}
\caption{Kinematic point used for $u\bar{u}\to G\to \gamma\gamma$.}\label{tab:ledpoint}
\end{table*}

\begin{table*}
\begin{center}
\small
\begin{tabular}{|lll|}
\hline
\multicolumn{3}{|c|}{\small Results for $u\bar{u}\to G\to \gamma\gamma$ with the kinematic point from Table~\ref{tab:ledpoint}.}\\
\hline
&\multicolumn{1}{l}{\small \GOSAM{} result} &
\multicolumn{1}{l|}{\small analytic result (Ref.~\cite{Kumar:2009nn})} \\
\hline
$a_{0}$ &  $\phantom{-}2.6456413225916027\times10^{-8}$ &
$\phantom{-}2.6456413225916010\times10^{-8}$ \\
$c_0/a_0$ & $\phantom{-}1.1594725347858084$  & $\phantom{-}1.1594725347858106$  \\
$c_{-1}/a_0$ & $-4.0000000000000009$ & $-4.0000000000000000$ \\
$c_{-2}/a_0$ & $-2.6666666666666661$ & $-2.6666666666666666$ \\
\hline
\end{tabular}
\end{center}
\caption{Result for the virtual amplitude 
including the QCD corrections to
 $u\bar{u}\to G\to \gamma\gamma$ within 
ADD models of large extra dimensions.}\label{tab:ledresult}
\end{table*}

\section{Conclusions}
\label{sec:conclusion}
We have presented the program package \GOSAM-2.0, which is a highly 
automated tool to calculate one-loop multi-particle amplitudes.
As the amplitudes at a first stage are produced in an algebraic form, 
the program offers a lot of flexibility concerning the 
particle content and the couplings, the choice of the reduction method 
and the treatment of the rational parts.

\GOSAM-2.0 can be used to calculate NLO QCD  corrections 
both {\em within and Beyond}
the Standard Model, as well as {\em electroweak} corrections, 
in combination with a Monte Carlo program providing the tree-level and NLO real radiation 
parts. 
The latter can be interfaced using the Binoth-Les-Houches-Accord, where both 
{\sc BLHA1} and {\sc BLHA2} standards are supported.
The automated interface to various Monte Carlo programs also offers the 
possibility to produce parton showered events and to compare different 
shower Monte Carlo event generators at NLO level.

We also note that the structure of the code is favourable to be used as a building block 
for the one-loop virtual times singly unresolved real radiation part entering NNLO calculations.

\GOSAM-2.0 contains many important new features.
The installation procedure is extremely simple: all dependencies 
are provided in one package, and an install script is 
building the whole package in a completely automated way.
Setting up a process is also very user-friendly: the user only 
has to fill out a well documented text file, the {\em process card}, 
where the program automatically chooses appropriate default values for 
unspecified options.

Improvements in the code generation compared to version 1.0 lead to more
compact and faster code. \GOSAM-2.0 also contains a new integrand
reduction method, the integrand decomposition via Laurent expansion,
implemented in the library \NINJA{}, which leads to a considerable
gain in stability and speed, in particular for amplitudes containing
internal masses.

The range of applicability of  \GOSAM{} also has been extended considerably. 
In particular, 
integrals where the rank exceeds the number of propagators (needed
e.g. in effective theories) are fully supported, and propagators for spin-2 particles are
implemented.  The complex mass scheme is supported, including the complexification
of the couplings, and several electroweak schemes can be chosen. 
Moreover, a new system for stability tests
and the rescue of `unstable' phase space points has been implemented. In addition,   
the program offers the possibility to produce spin-and colour correlated tree-level
matrix elements.
As a consequence, \GOSAM-2.0 can provide all the building blocks needed 
by modern Monte Carlo programs to construct a full NLO event generator, 
for QCD corrections both within and beyond the Standard Model, as well as 
electroweak corrections.

Therefore, 
to follow the strive for precision in the next phases of LHC data taking
as well as at a future Linear Collider, not only regarding QCD corrections, 
\GOSAM-2.0 can serve as a highly valuable tool.

\begin{acknowledgements}
We would like to thank the Herwig++ members J.~Bellm, S.~Gieseke, S.~Pl\"atzer, D.~Rauch 
and C.~Reuschle for fruitful interaction concerning the 
implementation of the BLHA2 interface.
We are grateful to S.~Pozzorini and C.~Papadopoulos for comparisons 
and to P.-F.~Monni for helpful discussions.
We also would like to thank Joscha~Reichel for collaboration on the 
BSM application  of GoSam involving spin-2 particles.
Finally, we are indebted to Thomas~Reiter for setting the groundwork 
GoSam is based on.
The work of G.C. was supported by DFG
Sonderforschungsbereich Transregio 9, Computergest\"utzte Theoretische Teilchenphysik.
P.M., H.v.D., G.L. and T.P. are supported by the Alexander von
Humboldt Foundation, in the framework of the Sofja Kovaleskaja Award Project
``Advanced Mathematical Methods for Particle Physics'', endowed by the German
Federal Ministry of Education and Research.
The work of G.O. was supported in part by the National Science Foundation
under Grant PHY-0855489 and PHY-1068550.
\end{acknowledgements}

\appendix
\renewcommand \thesection{\Alph{section}}
\renewcommand{\theequation}{\Alph{section}.\arabic{equation}}
\setcounter{equation}{0}

\section{Commented example of an input card}
\label{sec:appendix}
Here we give a commented example for the process 
$e^+e^-\rightarrow t\bar{t}$.

In the following it is assumed that the process
$e^+e^-\rightarrow t\bar{t}$ should be calculated to order
$\mathcal{O}(\alpha\alpha_s)$ (QCD corrections). We neglect
the exchange of a $Z$ or a Higgs boson and treat the electron as massless.
The output directory is assumed to be in the relative path
\texttt{eett}. 
A template file for a generic
process card (called {\tt eett.in} here) can be generated by invoking the shell command\\
{\tt gosam.py --template eett.in}\\
The template file {\tt eett.in} then should be edited by the user to define the process
specifications.
All lines starting with \# are comments.

At this point we would like to emphasize that almost all specifications 
in the process card are options, which will take default values if they are not 
filled in by the user. The paths to the libraries will be inserted 
automatically by the install script.
The only mandatory fields are the {\tt in} and {\tt out} 
particles, the perturbative order and the path where to store the process files.
Therefore, a minimal process card can look like this:
\begin{lstlisting}[gobble=3,%
     numbers=left,caption={{\tt eett.in}},%
     basicstyle=\ttfamily]
1  process_path=eett
2  in=    e+, e-
3  out=   t, t~
4  order= gs, 0, 2
\end{lstlisting}

In order to populate the process subdirectory specified under {\tt process\_path}
with files for code generation 
one invokes\\
{\tt gosam.py eett.in}\\
This will create the subdirectory structure described in Section \ref{sec:usage}.

\medskip

In the following, we will give detailed comments to all the fields and options available 
in the process card for the example {\tt eett.in}. 
(Please note that the line numbers on the left
are only included for better readability and should \emph{not} be included
in your input file).
\begin{lstlisting}[gobble=3,%
     numbers=left,caption={{\tt eett.in}},%
     basicstyle=\ttfamily]
 1 process_name=eett
 2 process_path=eett
 3 in=    e+, e-
 4 out=   t, t~
 5 model= smdiag
 6 model.options=ewchoose
 7 order= gs, 0, 2
 8 zero=me
 9 one=gs,e
10 regularisation_scheme=dred
11 helicities=
12 qgraf.options=onshell,notadpole,nosnail
13 qgraf.verbatim= True=iprop[Z, 0, 0];\n\
14                 true=iprop[H, 0, 0];
15 qgraf.verbatim.lo=
16 qgraf.verbatim.nlo=
17 polvec=numerical
18 diagsum=True
19 reduction_programs=ninja,golem95,samurai
20 extensions=shared
21 debug=nlo
22 select.lo=
23 select.nlo=
24 filter.lo=
25 filter.nlo=
26 filter.module=
27 renorm_beta=True
28 renorm_mqwf=True
29 renorm_decoupling=True
30 renorm_mqse=True
31 renorm_logs=True
32 renorm_gamma5=True
33 reduction_interoperation=-1
34 reduction_interoperation_rescue=-1
35 samurai_scalar=2
36 nlo_prefactors=0
37 PSP_check=True
38 PSP_rescue=True
39 PSP_verbosity=False
40 PSP_chk_th1=8
41 PSP_chk_th2=3
42 PSP_chk_th3=5
43 PSP_chk_kfactor=10000
44 reference-vectors=
45 abbrev.limit=0
46 templates=
47 qgraf.bin=qgraf
48 form.bin=form
49 form.threads=2
50 form.tempdir=/tmp
51 haggies.bin=
52 fc.bin=/usr/bin/gfortran
53 python.bin=python
54 ninja.fcflags=
55 ninja.ldflags=
56 samurai.fcflags=
57 samurai.ldflags=
58 golem95.fcflags=
59 golem95.ldflags=
60 r2=explicit
61 symmetries=family,generation
62 crossings=
\end{lstlisting}

The comments to the file {\tt eett.in} are as follows. 

\begin{enumerate}
\item[1] Setting a process name is optional but recommended. All module
names will be prefixed with the  process name (e.g. \texttt{precision}
$\to$ \texttt{eett\_precision}). This will avoid name conflicts if at
a later stage more than one matrix elements are linked into one
executable.
\item[2] The item \texttt{process\_path} specifies the directory
to which all generated files and directories are written.
Specification of a process path is mandatory.
\item[3--4] The items \texttt{in} and \texttt{out} specify the particles
of the initial and final state. The particle names must be defined in the
selected model file. As the model files usually define mnemonics for the
particle names there might be several ways of specifying the same process.
Instead of `\lstinline[basicstyle=\ttfamily]{e+}'
one could have written `\lstinline[basicstyle=\ttfamily]{ep}'
or `\lstinline[basicstyle=\ttfamily]{positron}'.
For a complete list of
alternative particle names please refer to the documentation of the
corresponding model file.
\\Specifying \texttt{in} and \texttt{out} particles is mandatory.
\item[5] The option \texttt{model} specifies which model files should
be used in order to generate and evaluate the diagrams. 
How to import models in {\tt UFO} or {\tt LanHep} format is described in 
Section \ref{sec:model}.
The default for this field is {\tt smdiag}, i.e. the built-in Standard Model file
with a diagonal CKM matrix.
\item[6] The option \texttt{model.options} can be used to pass options 
which are specific to a certain model. The default is {\tt ewchoose}, 
which means that the electro-weak scheme is selected automatically according to the 
given input parameters.
\item[7] The item \texttt{order} is a comma separated list with
three entries. The first entry specifies a symbol that denotes a coupling
constant. In the Standard Model file \texttt{sm} the only two possibilities
are `\lstinline[basicstyle=\ttfamily]{gs}' for the strong coupling constant
$g_s$ and `\lstinline[basicstyle=\ttfamily]{e}' for the electroweak coupling.
The second number is the power of the chosen coupling constant for the
tree-level diagrams and the third number specifies the power of that
coupling constant for the one-loop diagrams.
Note that the numbers
refer to the powers in the diagrams of the amplitude
rather than the squared amplitude. In the above example the
string `\lstinline[basicstyle=\ttfamily]{gs, 0, 2}' specifies that
the tree-level diagrams should be of order $g_s^0$ and the one-loop
diagrams should be of order $g_s^2$ and an unspecified
power of $e$ in both cases. 
If there is no tree level, i.e. the process is loop induced, 
the keyword \texttt{NONE} should be put as second item in the list,
instead of the tree level power of the coupling.\\
The values of {\tt order} are
translated into a \texttt{vsum} constraint in the file \texttt{qgraf.dat}.
\\This field is mandatory.
\item[8--9] The keywords \texttt{zero} and \texttt{one} specify
a set of symbols that should be treated as zero (resp. one). These
simplifications are applied at the symbolical level. Only symbols
that appear in the \FORM{} interface of the model file should be
specified here (masses, couplings, CKM-matrix elements, etc).
In the example we specify the electron mass
`\lstinline[basicstyle=\ttfamily]{me}' to be zero and we do not keep
the coupling constants in the calculation explicitly ($g_s=e=1$).
\\These options can be omitted.
\item[10] The option \texttt{regularisation\_scheme} 
allows to choose the dimensional regularisation scheme, in our example {\tt dred} for 
dimensional reduction, which is the default.
{\tt cdr} for ``conventional dimensional regularisation" is also possible.
\item[11] \texttt{helicites}:  a comma separated list of helicities to be calculated. 
An empty list means that all possible helicities should be generated.         
The characters correspond to particles 1, 2, ... from left to right.   \\
 Example:                                                       
 $e^+ e^- \to \gamma \gamma$:    \\                               
 Only three helicity configurations are required; the other ones are         
  either zero or can be obtained by symmetry                 
  transformations. This corresponds to\\                                       
      {\tt  helicities=+-++,+-+-,+---}  \\
 Multiple helicities can be encoded in patterns, which are expanded
    at the time of code generation. For more details we refer to the manual.      
\item[12]  \texttt{qgraf.options=onshell,notadpole,nosnail}: 
a list of options which is passed to \QGRAF{} via the 'options' line.
Possible values (as of qgraf.3.1.1) are the following keywords:
 onepi, onshell, nosigma, nosnail, notadpole, floop, topol.        
In our example, it means that external lines are on-shell, i.e. 
do not contain selfenergy corrections, and that tadpole and snail diagrams are discarded.    
We refer to the \QGRAF{} documentation for more details.
\item[13-16] 
 The value of the option \texttt{qgraf.verbatim} is
passed verbatim to the file \texttt{qgraf.dat}.
In our example we suppress the generation of diagrams containing Higgs and $Z$ bosons.
As these commands are passed verbatim to \QGRAF, no mnemonic names
are allowed here, e.g. the Higgs particle has to be denoted by
`\lstinline[basicstyle=\ttfamily]{H}' and cannot be replaced by
`\lstinline[basicstyle=\ttfamily]{h}'.
For a complete list of available options, please consult the
\QGRAF{} manual. For a complete list of particle names we refer to the 
documentation of the corresponding model file.
\\These options can be omitted.
\item[17] \texttt{polvec}: by default ({\tt polvec=numerical}), 
numerical polarisation vectors are used for the massless gauge bosons, 
rather than producing separate code for each helicity
(see Section \ref{sec:numpolvec}). 
To switch off the use of numerical polarisation vectors, use {\tt polvec=explicit}.
\item[18] \texttt{diagsum}:  if {\tt True}, one-loop diagrams sharing some propagators
are combined before  the algebraic reduction.
The default is {\tt diagsum = True}. 
\item[19] The option \texttt{reduction\_programs} 
allows to choose the amplitude reduction method. If several choices are given, 
the code is produced such that the reduction methods can be switched at runtime.
The default is {\tt ninja, golem95}.
\item[20] {\tt extensions}: 
this option  contains a list of useful extensions
to the core of the program, which operate at the code generation stage. 
The currently available extensions are 
\begin{itemize}
\item {\tt autotools}:  use autotools to generate Makefiles            
\item {\tt shared}: create shared libraries (i.e. dynamically linkable code
rather than static libraries). This extension is 
enabled by default when using the {\tt autotools} extension.    
\item {\tt f77}: in combination with the BLHA interface it generates
                     a file {\tt olp\_module.f90} linkable with Fortran77.      
\item {\tt noformopt}:  disables diagram optimization using \FORM        
\item {\tt gaugecheck}:  modifies the massless gauge boson wave functions to allow for
a check of gauge invariance for processes involving 
gluons or photons. 
\item {\tt customspin2prop} allows to replace the propagator of spin-2 particles  
with a custom function (we refer to the manual for details).
\end{itemize}
In our example {\tt shared} tells the program to 
build dynamic rather than static libraries.
\item[21] \texttt{debug}: can take the values  {\tt lo}, {\tt nlo}, {\tt all}.
It sets the level of information printed to the file {\tt matrix/debug.xml} when running the test program.
\item[22] \texttt{select.lo}: can be used to select/discard diagrams by their diagram numbers.
It can contain a list of integer numbers, indicating leading order diagrams to be
selected. If no list is given, all diagrams are selected.       
Otherwise, all diagrams whose numbers are not in the list will be discarded.          
The list may also contain ranges, with increments different from one, e.g.                                    
{\tt select.lo=1,2,5:10:3, 50:53} is equivalent to {\tt select.lo=1,2,5,8,50,51,52,53}, 
i.e. the 3 in    {\tt 5:10:3} is the increment.
\item[23] \texttt{select.nlo}: analogous to \texttt{select.lo},  for the one-loop diagrams.
\item[24] \texttt{filter.lo}:  a python function which provides a filter for tree diagrams.    \\
Example:    {\tt filter.lo=lambda d: d.iprop(Z) == 1  and d.vertices(Z, U, Ubar) == 0 } filters out
diagrams containing exactly one $Z$ propagator and no $Zu\bar{u}$ couplings. 
\item[25] \texttt{filter.nlo}:  analogous to \texttt{filter.lo},  for the one-loop diagrams.
For details we refer to the manual.
\item[26] \texttt{filter.module}: a python file of predefined functions which can be used as filters.   
\item[27] \texttt{renorm\_beta}: activates or disables beta function renormalisation. The default is  {\tt True}. 
\item[28] \texttt{renorm\_mqwf}:  activates or disables UV countertems coming from
external massive quarks. The default is  {\tt True}.
\item[29] \texttt{renorm\_decoupling}:   activates or disables UV counterterms coming from  
massive quark loops. The default is  {\tt True}.
\item[30] \texttt{renorm\_mqse}:  activates or disables the UV counterterm coming from the 
massive quark propagators. The default is  {\tt True}.
\item[31] \texttt{renorm\_logs}: activates or disables the logarithmic finite terms  
associated with the UV counterterms.   The default is  {\tt True}.
\item[32] \texttt{renorm\_gamma5}: activates finite renormalisation for axial couplings in the 
 \tHV{} scheme (CDR).                      
Implemented for QCD only, works only with the built-in model files.    The default is  {\tt True}.
\item[33] \texttt{reduction\_interoperation}: denotes the reductuion libraries to be used. 
  Possible values are: ninja, samurai, golem95 (listing all of them simultaneously is possible). 
  A value of -1 lets \GOSAM{} decide.
  See {\tt common/config.f90} for details.                              
\item[34] \texttt{reduction\_interoperation\_rescue}: specifies the reduction library to be used to rescue `unstable points'. A value of -1 lets \GOSAM{} decide.
\item[35] \texttt{samurai\_scalar}: integer which specifies the library \SAMURAI{} 
chooses for the basis integrals. 1: {\sc QCDLoop}, 2: {\sc OneLOop}, 3: \GOLEMVC. The default is 2. 
\item[36] \texttt{nlo\_prefactors}:  can take the integer values 0,1,2, which have the        
    following meaning:    
    \begin{enumerate}                                          
    \item[0]: a factor of $\alpha_{(s)}/(2\pi)$ is not included in the NLO result
    \item[1]: a factor of $1/(8\pi^2)$ is not included in the NLO result       
    \item[2]: the NLO result includes all prefactors   (see also manual).
    \end{enumerate}                           
    Note, however, that the factor of $1/\Gamma(1-\epsilon)$ is not         
    included in any of the cases.                                   
    Please note also that {\tt nlo\_prefactors=0} is enforced in {\tt test.f90}
    in order to recognize rational numbers for the 
    pole coefficients. In the OLP interface mode (BLHA/BLHA2), the default is
    {\tt nlo\_prefactors=2}.                                 
\item[37] \texttt{PSP\_check}: allows to switch the stability test of the full amplitude for 
each phase space point on or off.
If \texttt{PSP\_check} is set to {\tt False}, the following flags concerning 
\texttt{PSP\_rescue} and the various thresholds for the rescue system have no effect.
Details about the stability tests are given in Section \ref{sec:rescue}.
Please note that this test only works for QCD  with the built-in model files.
The default is \texttt{PSP\_check= True}. 
\item[38] \texttt{PSP\_rescue}:  
    activates the phase space point rescue system based on the estimated       
    accuracy of the finite part.                                    
    The accuracy is estimated using information on the single       
    pole accuracy and the cancellation between the cut-constructible    
    part and $R_2$.                                                    
The default is \texttt{PSP\_rescue= True}. 
\item[39] \texttt{PSP\_verbosity}:     sets the verbosity of the {\tt PSP\_check}. 
{\tt verbosity = False} means no output,                                     
{\tt verbosity = True} means that bad points are written to a file \\{\tt gs\_badpts.log}.
 The default is {\tt verbosity = False}.
\item[40] \texttt{PSP\_chk\_th1}: an integer indicating the  number of desired accurate digits
of the single pole coefficient. For poles coefficients more precise than this
threshold the finite part is not checked separately.
Note that this works only for QCD, with the built-in model files.    
The default is 8. 
\item[41] \texttt{PSP\_chk\_th2}: threshold (number of accurate digits) to declare a 
phase space point  as {\it bad point}, based on the precision of the pole coefficient.
Points with precision less than this threshold are directly reprocessed with 
    the rescue system (if available), or declared as unstable. According to the
    verbosity level set, such points are written to a file and not used when
    the code is interfaced to an external Monte Carlo program using the new BLHA2 standards.
The default is 3. 

\item[42] \texttt{PSP\_chk\_th3}:  threshold (number of accurate digits) to declare a 
phase space point  as {\it bad point}, based on the precision of 
    the finite part estimated with a {\it rotation}. According to the     
    verbosity level set, such points are written to a file and not  
    used when the code is interfaced to an external Monte Carlo program 
    using the new BLHA2 standards.           
 The default is 5. 
\item[43] \texttt{PSP\_chk\_kfactor}: threshold on the K-factor to declare a 
phase space point  as {\it bad point}. According 
to the verbosity level set, such points are written to a file and 
    not used when the code is interfaced to an external Monte Carlo program 
    using the new BLHA2 standards.                    
The default is 10000.
\item[44] \texttt{reference-vectors}: comma separated list of  reference vectors for massive 
fermions and vector bosons.
If no reference vectors are assigned here, the program picks the reference vectors automatically. 
Each entry of the list has to be of the form $\langle index\rangle :\langle index\rangle $.  
Example: \\
{\tt in=g,u}\\                                                       
{\tt out=t,W+ }\\                                                       
{\tt reference-vectors=1:2,3:4,4:3 }\\                                  
    In this example, the gluon (particle 1) takes the momentum $k_2$   
    as reference momentum for the polarisation vector. The massive  
    top quark (particle 3) uses the light-cone projection $l_4$ of the
    W-boson as reference direction for its own momentum splitting.  
    Similarly, the momentum of the W-boson is split into a direction
    $l_4$ and one along $l_3$.                                            
\item[45] \texttt{abbrev.limit}: maximum number of instructions per subroutine when calculating  
 abbreviations. The default is 0, which means that no maximum is set.
\item[46] \texttt{templates}:   path pointing to the directory containing the template 
    files for the process. If not set, the program uses the directory      
    $\langle$ gosam\_path$\rangle$/templates.                                         
    The directory must contain a file called {\tt template.xml}.
\item[47] \texttt{qgraf.bin}: path to the {\tt QGraf} executable.  
The default path will be set by the installation script.
\item[48] \texttt{form.bin}:  path to the \FORM{} executable.  
The default path will be set by the installation script.
\item[49] \texttt{form.threads}: the number of \FORM{} threads when using {\tt tform}, 
the parallel version of \FORM.  The default is 2.
\item[50] \texttt{form.tempdir}: the temporary directory where \FORM{} 
can store (large) intermediate files.  the default is {\tt /tmp}.
\item[51] \texttt{haggies.bin}: path to the {\tt haggies} executable.
The default path will be set by the installation script.
\item[52] \texttt{fc.bin}: path to the {\tt Fortran} compiler. 
The default path will be set by the installation script.
\item[53] \texttt{python.bin}:  path to the {\tt python} executable. 
The default path will be set by the installation script.
\item[54] \texttt{ninja.fcflags}: compiler flags to compile with \NINJA. 
The default will be set by the installation script.
\item[55] \texttt{ninja.ldflags}: {\sc ldflags} required to link the \NINJA{} library.    
The default will be set by the installation script.
\item[56] \texttt{samurai.fcflags}:  compiler flags to compile with \SAMURAI. 
The default will be set by the installation script.
\item[57] \texttt{samurai.ldflags}: {\sc ldflags} required to link the \SAMURAI{} library.    
The default will be set by the installation script. 
\item[58] \texttt{golem95.fcflags}: compiler flags to compile with \GOLEMVC{}. 
The default will be set by the installation script. 
\item[59] \texttt{golem95.ldflags}:  {\sc ldflags} required to link the \GOLEMVC{} library.    
The default will be set by the installation script. 
\item[60] \texttt{r2}: treatment of the rational part $R_2$. The possibilities are:
\begin{itemize}
\item {\tt implicit}: $\mu^2$ terms are kept in the numerator and reduced at runtime,
\item {\tt explicit}: $\mu^2$ terms are reduced analytically, 
\item {\tt off}: all $\mu^2$ terms are set to zero.
\end{itemize}
 The default is {\tt r2=explicit}.
\item[61] \texttt{symmetries}: this information is used when the list of helicity configurations is generated. 
An empty list means that all helicity configurations will be generated, even if some of them could be 
mapped onto each other.
Possible values are:  
\begin{itemize}
\item {\tt flavour}:  assumes that no flavour changing interactions are present.             
When calculating the list of helicities, fermions
with PDG codess 1-6 are assumed not to mix.                         
\item {\tt family}:  flavour changing only within families.           
             When calculating the list of helicities, fermion lines
        with PDG codes 1-6 are assumed to mix only within families,        
        i.e. a quark line connecting an up with a down quark would   
        be considered, while up-bottom would be discarded.                      
\item  {\tt lepton}:   means for leptons what `flavour' means for quarks.
\item  {\tt generation}: means for leptons what `family' means for quarks.
\item $\langle n\rangle =\langle h\rangle $:    restriction of particle helicities,             
             e.g. {\tt 1=-, 2=+} specifies helicities of particles 1 and 2.
\item \%$\langle n\rangle =\langle h\rangle $   restriction by PDG code,                        
 e.g. \%23=+- specifies the helicity of all Z-bosons to be '+' and '-' only (no '0' polarisation),  \\              
             \%$\langle n\rangle $ refers to both $+n$ and $-n$,    \\                      
             \%+$\langle n\rangle $ refers to $+n$ only,                                 
             \%-$\langle n\rangle $ refers to $-n$ only.  
\end{itemize}	                                   
\item[62] \texttt{crossings}:  a list of crossed processes derived from this process.          
    For each process in the list a module similar to {\tt matrix.f90} is  generated.  \\
    Example: \\                                                 
    {\tt process\_name=ddx\_uux }\\                                           
    {\tt in=1,-1 }\\                                                          
    {\tt out=2,-2  }\\                                                       
    {\tt crossings = dxd\_uux: -1 1 $\to$ 2 -2, ud\_ud: 2 1 $\to$ 2 1  }    
\end{enumerate}

\section{Higher rank integrals}
\label{sec:appendixB}

Higher rank integrals are implemented in all reduction libraries included in \GOSAM{}.
\NINJA{} and \SAMURAI{} are based on integrand reduction, as described in Section~\ref{sec:ninja}, 
\GOLEMVC{} provides tensor integrals, using a tensor reduction method and a basis of scalar integrals which 
has been designed to provide numerical stability in problematic phase space regions, 
for example in the limit of small Gram determinants.

In the following we briefly sketch the main features of the higher rank extensions 
for both approaches, more details can be found in \cite{vanDeurzen:2013pja,Guillet:2013msa,Mastrolia:2012bu}.

\subsection{Integrand reduction approach}

If the rank $r$ of a one-loop integrand is not larger than the number of propagators $N$, the
respective integral can be written as the following combination of
known master integrals
\begin{align}
  {\mathcal{M}} = {}&
\sum_{\{i_1, i_2, i_3, i_4\}}\bigg\{
          c_{0}^{ (i_1 i_2 i_3 i_4)} I_{i_1 i_2 i_3 i_4} +  
          c_{4}^{ (i_1 i_2 i_3 i_4)} I_{i_1 i_2 i_3 i_4}[\mu^4] 
\bigg\} \pagebreak[1] \nonumber \\
     & +
\sum_{\{i_1, i_2, i_3\}}\bigg\{
          c_{0}^{ (i_1 i_2 i_3)} I_{i_1 i_2 i_3} +
          c_{7}^{ (i_1 i_2 i_3)} I_{i_1 i_2 i_3}[\mu^2]
\bigg\} \pagebreak[1] \nonumber \\
     & +
\sum_{\{i_1, i_2\}}\bigg\{
          c_{0}^{ (i_1 i_2)} I_{i_1 i_2} 
        + c_{1}^{ (i_1 i_2)} I_{i_1 i_2}[(q+p_{i_1})\cdot e_2 ] \nonumber \\
        & \qquad
        + c_{2}^{ (i_1 i_2)} I_{i_1 i_2}[((q+p_{i_1})\cdot e_2)^2 ]         +  c_{9}^{ (i_1 i_2)} I_{i_1 i_2}[\mu^2]  \bigg\} \pagebreak[1] \nonumber \\ 
& + 
\sum_{i_1}
      c_{0}^{ (i_1)} I_{i_1},   \label{eq:integraldecomposition}
\end{align}
where
\begin{align}
  I_{i_1 \cdots i_k}[\alpha] & \equiv \int d^{4-2\epsilon} q {\alpha \over D_{i_1} \cdots D_{i_k} }, 
\qquad
I_{i_1 \cdots i_k} \equiv   I_{i_1 \cdots i_k}[1], \nonumber \\
  D_j & = (q+p_j)^2-m_j^2
\end{align}
with
\begin{eqnarray}
I_{i_1 i_2 i_3 i_4}[\mu^4] &=& -\frac{1}{6} +{\mathcal{O}}(\epsilon)\nn \\
I_{i_1 i_2 i_3}[\mu^2]     &=&  \frac{1}{2} +{\mathcal{O}}(\epsilon) \nn\\
I_{i_1 i_2}[\mu^2]         &=&
-\frac{1}{6}\left(p_{i_1}^2-3(m_{i_1}+m_{i_2}) \right)+{\mathcal{O}}(\epsilon)\;.
\end{eqnarray}
In the case where $r=N+1$ the integral is generalized as
\begin{align}
  {\mathcal{M}}^{(r=N+1)} = {}& {\mathcal{M}}^{(r=N)}+ 
\sum_{\{i_1, i_2, i_3\}}\, 
          c_{14}^{ (i_1 i_2 i_3)} I_{i_1 i_2 i_3}[\mu^4]
\pagebreak[1] \nonumber \\
     & +
\sum_{\{i_1, i_2\}}\bigg\{
          c_{10}^{ (i_1 i_2)} I_{i_1 i_2}[\mu^2\, (q+p_{i_1})\cdot e_2) ]
        + c_{13}^{ (i_1 i_2)} I_{i_1 i_2}[((q+p_{i_1})\cdot e_2)^3 ] \bigg\} \pagebreak[1] \nonumber \\ 
& + 
\sum_{i_1}
     \bigg\{ c_{14}\, I_{i_1}[\mu^2] + c_{15}^{ (i_1)}\, I_{i_1}[((q+p_{i_1})\cdot e_3)((q+p_{i_1})\cdot e_4)] \bigg\}  \label{eq:hrintegraldecomposition}.
\end{align}

The three integrals in Eq.~\eqref{eq:hrintegraldecomposition} whose
numerator is proportional to $\mu^2$ are finite and contribute to the
rational part of the amplitude.  They have been computed in
Ref.~\cite{Reiter:2009kb,Mastrolia:2012bu} and they read
\begin{align}
  I_{i_1 i_2 i_3}[\mu^4] ={}& \frac{i \pi^2}{6}\left( \frac{s_{i_2 i_1} + s_{i_3 i_2} + s_{i_1 i_3}}{4} - m_{i_1}^2 - m_{i_2}^2 - m_{i_3}^2 \right) +{\mathcal{O}}(\epsilon)\\
  I_{i_1 i_2}[\mu^2\, ((q+p_{i_1}) \cdot e_2)] ={}& i \pi^2\, \frac{((p_{i_2}-p_{i_1}) \cdot e_2)}{12}\Big( s_{i_2 i_1} - 2\, m_{i_1}^2 - 4\, m_{i_2}^2 \Big)  +{\mathcal{O}}(\epsilon) \\
  I_{i_1}[\mu^2] ={}& \frac{i \pi^2\, m_{i_1}^4}{2}   +{\mathcal{O}}(\epsilon)
\end{align}
where $s_{ij}\equiv (p_i-p_j)^2$.  The tadpole of rank 2 appearing in
Eq.~\eqref{eq:hrintegraldecomposition} can be written in terms of the
scalar tadpole $I_{i_1}$ as
\begin{equation}
  I_{i_1}[((q+p_{i_1})\cdot e_3)\, ((q+p_{i_1})\cdot e_4)] ={} m_{i_1}^2\, \frac{(e_3\cdot e_4)}{4}\left( I_{i_1} + \frac{i \pi^2\, m_{i_1}^2}{2} \right)   + {\mathcal{O}}(\epsilon).
\end{equation}
Finally, since the vector $e_{2}$ can always be chosen to be massless,
the bubble integral of rank 3 appearing in
Eq.~\eqref{eq:hrintegraldecomposition} is proportional to the form
factor $B_{111}$,
\begin{equation}
  I_{i_1 i_2}[((q+p_{i_1})\cdot e_2)^3 ] = ((p_{i_2}-p_{i_1})\cdot e_2)^3\, B_{111}(s_{i_2 i_1},m_{i_1}^2, m_{i_2}^2).
\end{equation}
The latter can be computed using the formulas of
Ref.~\cite{Stuart:1987tt}, as a function of form factors of scalar
integrals $B_0$.  In the special case with $s_{i_2 i_1} = 0$ we use
Eq.~(A.6.2) and (A.6.3) of that reference.  For the general case
$s_{i_2 i_1} \neq 0$ we use instead the following
formula~\cite{Peraro:2014cba}
\begin{align}
  B_{111}(s_{i_2 i_1},m_{i_1}^2,m_{i_2}^2) ={}\frac{1}{4\, s_{i_2 i_1}^3} \bigg\{& s_{i_2 i_1}\, \Big(m_{i_1}^2 
\, I_{i_1}+ I_{i_1}[\mu^2]-m_{i_2}^2 
\, I_{i_2}-I_{i_2}[\mu^2] \nonumber \\ & - 4 
\, I_{i_1 i_2}[\mu^2\, ((q+p_{i_1})\cdot (p_{i_2}-p_{i_1}))] \nonumber \\ & - 4 \, m_{i_1}^2
\, I_{i_1 i_2}[(q+p_{i_1})\cdot (p_{i_2}-p_{i_1})]\Big) \nonumber \\
& + 4\, (m_{i_2}^2-m_{i_1}^2-s_{i_2 i_1}) \, I_{i_1 i_2}[((q+p_{i_1})\cdot (p_{i_2}-p_{i_1}))^2] \bigg\}.
\end{align}

\subsection{Tensor reduction approach}

In the tensor reduction approach, the tensor integrals are written in terms of 
linear combinations of scalar {\it form factors} and all possible combinations of 
external momenta and metric tensors carrying the Lorentz structure. 
The form factors themselves are then reduced to a convenient set of basis integrals. 
It is well known that, due to the 4-dimensionality (resp. $D=4-2\epsilon$ dimensionality in dimensional regularisation) 
of space-time, integrals with $N\geq 6$ can be reduced iteratively to 
5-point integrals.
Therefore form factors for $N\geq 6$ are never needed.
The general form factor decomposition of an arbitrary tensor integral can be written as
\begin{eqnarray}\label{eq:formfactordef}
I^{D,\mu_1\ldots\mu_r}_N(a_1,\ldots, a_r; S)&=&
\int\!\!\frac{\diff[D]k}{i\pi^{D/2}}\frac{
q_{a_1}^{\mu_1} \ldots q_{a_r}^{\mu_r}}{
\prod_{j=1}^N(q_j^2-m_j^2+i\delta)}\\
&=&
\sum_{j_1,\ldots,j_r\in S}
   \left[\Delta_{j_1\cdot}^{\cdot}\cdots\Delta_{j_r\cdot}^{\cdot}%
   \right]^{\{\mu_1\ldots\mu_r\}}_{\{a_1\ldots a_r\}}
   A^{N,r}_{j_1\ldots j_r}(S)\nn\\
&&+ \sum_{j_1,\ldots,j_{r-2}\in S}
   \left[g^{\cdot\cdot}%
   \Delta_{j_1\cdot}^{\cdot}\cdots\Delta_{j_{r-2}\cdot}^{\cdot}%
   \right]^{\{\mu_1\ldots\mu_r\}}_{\{a_1\ldots a_r\}}
   B^{N,r}_{j_1\ldots j_{r-2}}(S)
\nn\\
&&+ \sum_{j_1,\ldots,j_{r-4}\in S}
   \left[g^{\cdot\cdot}g^{\cdot\cdot}%
   \Delta_{j_1\cdot}^{\cdot}\cdots\Delta_{j_{r-4}\cdot}^{\cdot}%
   \right]^{\{\mu_1\ldots\mu_r\}}_{\{a_1\ldots a_r\}}
   C^{N,r}_{j_1\ldots j_{r-4}}(S)
\nn\\
&&+ \sum_{j_1,\ldots,j_{r-4}\in S}
   \left[g^{\cdot\cdot}g^{\cdot\cdot}g^{\cdot\cdot}%
   \Delta_{j_1\cdot}^{\cdot}\cdots\Delta_{j_{r-6}\cdot}^{\cdot}%
   \right]^{\{\mu_1\ldots\mu_r\}}_{\{a_1\ldots a_r\}}
   D^{N,r}_{j_1\ldots j_{r-6}}(S)
\nn\\
&&+\;\ldots   \;,\nn
\end{eqnarray}
where $\Delta_{ij}^\mu=
r_i^\mu - r_j^\mu$ are differences of external momenta $r$, and $q_a=k+r_a$.
The notation $[\cdots]^{\{\mu_1\cdots\mu_r\}}_{\{a_1\cdots a_r\}}$ 
stands for the distribution of the $r$ Lorentz indices $\mu_i$, and the momentum 
labels $a_i$,  to the vectors $\Delta_{j\,a_i}^{\mu_i}$ and metric tensors 
in all distinguishable ways. 
Note that the choice $r_N=0$, $a_i=N \,\,\forall \,i$ leads to the well known representation 
in terms of external momenta where the labels $a_i$ are not necessary, 
but we prefer a completely shift invariant notation here.

$S$ denotes an ordered 
set of propagator labels, corresponding to the momenta forming 
the kinematic matrix ${\cal S}$, defined by 
\bea
\calst_{ij} &=&  (r_i-r_j)^2-m_i^2-m_j^2\;, \;\quad i,j\in\{1,\ldots,N\}\;.
\label{eqDEFS}
\eea

We should point out that the form factors of type $D^{N,r}_{j_1\ldots j_{r-6}}$ and beyond,
i.e. form factors associated with three or more metric tensors, 
are not needed for integrals where the rank $r$ does not exceed the number $N$ 
of propagators, no matter what the value of $N$ is,  
because integrals with $N\geq 6$ can be reduced algebraically to pentagons.

The program \GOLEMVC{} reduces the form factors $A,\ldots ,D$ internally to a set of 
basis integrals, i.e. the endpoints of the reduction 
(they do not form a basis in the mathematical sense, 
as some of them are not independent).
The choice of the basis integrals can have important effects on the numerical stability 
in certain kinematic regions.
Our reduction endpoints are 
4-point functions in 6 dimensions
$I_4^6$, which are IR and UV finite, 4-point functions in
$D+4$ dimensions, and various 3-point, 2-point and 1-point functions.
A special feature of \GOLEMVC{} is that the algebraic reduction 
to scalar basis integrals is automatically replaced by a stable and fast 
one-dimensional numerical integration of parametric integrals 
corresponding to tensor rather than scalar integrals 
in kinematic situations where a further reduction would lead to 
spurious inverse Gram determinants tending to zero. 
This leads to improved numerical stability. 

The extension of \GOLEMVC{} to higher rank integrals \cite{Guillet:2013msa}
follows the reduction formalism as outlined in \cite{Binoth:2005ff}. 
However, the extension of the formalism to rank six pentagons 
required some care, as the latter develop an UV divergence, 
and therefore ${\cal O}(\epsilon)$ terms occurring in the reduction 
need to be taken into account at intermediate stages.

The rational parts of all the integrals contained in \GOLEMVC{}
can be extracted separately, and analytic formulae for $r\leq N$ are provided in 
\cite{Binoth:2006hk}.
The results for those integrals which are relevant for the 
higher rank extension can be extracted from~\cite{Reiter:2009kb}, 
where formulae for all possible rational parts
are given in a general form.
The ones which are relevant for the higher rank extension which have not been 
given above already are listed explicitly here, where the notation conventions are
$k_{(D)}^{\mu}  =  \hat{k}_{(4)}^{\mu} + \tilde{k}_{(-2\eps)}^{\mu}, 
k_{(D)}^2  =  \hat{k}^2 + \tilde{k}^2$\,,
\begin{equation}
\label{eq:ktilde}
I_N^{D,\alpha;\mu_1\ldots\mu_r}(a_1,\ldots a_r; S)\equiv
\int\!\!\frac{\diff[D]k}{i\pi^{D/2}}\frac{\left(\tilde{k}^2\right)^\alpha
\hat{q}_{a_1}^{\mu_1}\cdots \hat{q}_{a_r}^{\mu_r}}{
\prod_{j=1}^N(q_j^2-m_j^2+i\delta)}\text{,}
\end{equation}
with the results~\cite{Guillet:2013msa}
\begin{eqnarray}
I_5^{D,3}(S)&=&-\frac{1}{12}+{\cal O}(\eps)\;,\\
I_5^{D,2;\mu_1 \mu_2}(a_1,a_2; S)&=&
-\frac{1}{48}\,g^{\mu_1\mu_2}+{\cal O}(\eps)\;,\nn\\
I_5^{D,1;\mu_1\cdots \mu_4}(a_1,\ldots,a_4; S)&=&-\frac{1}{96}\,
\left[g^{\mu_1\mu_2}g^{\mu_3\mu_4}+ g^{\mu_1\mu_3}g^{\mu_2\mu_4}
+g^{\mu_1\mu_4}g^{\mu_2\mu_3}\right]+{\cal O}(\eps)\;,\nn\\
\eps I_4^{D+6}(S)&=&\frac{1}{240}\left(\sum_{i,j=1}^4 (\Delta^2_{ij}-m_i^2-m_j^2)-2\sum_{i=1}^4 m_i^2\right)
+{\cal O}(\eps)\;.\nn
\end{eqnarray}


\bibliographystyle{JHEP}


\begin{thebibliography}{10}

\bibitem{Aad:2012tfa}
{\bf ATLAS} Collaboration, G.~Aad et~al., {\it {Observation of a new particle
  in the search for the Standard Model Higgs boson with the ATLAS detector at
  the LHC}},  {\em Phys.Lett.} {\bf B716} (2012) 1--29,
  [\href{http://xxx.lanl.gov/abs/1207.7214}{{\tt arXiv:1207.7214}}].

\bibitem{Chatrchyan:2012ufa}
{\bf CMS} Collaboration, S.~Chatrchyan et~al., {\it {Observation of a new boson
  at a mass of 125 GeV with the CMS experiment at the LHC}},  {\em Phys.Lett.}
  {\bf B716} (2012) 30--61, [\href{http://xxx.lanl.gov/abs/1207.7235}{{\tt
  arXiv:1207.7235}}].

\bibitem{Hahn:2010zi}
T.~Hahn, {\it {Feynman Diagram Calculations with FeynArts, FormCalc, and
  LoopTools}},  {\em PoS} {\bf ACAT2010} (2010) 078,
  [\href{http://xxx.lanl.gov/abs/1006.2231}{{\tt arXiv:1006.2231}}].

\bibitem{Bevilacqua:2011xh}
G.~Bevilacqua, M.~Czakon, M.~Garzelli, A.~van Hameren, A.~Kardos, et~al., {\it
  {HELAC-NLO}},  {\em Comput.Phys.Commun.} {\bf 184} (2013) 986--997,
  [\href{http://xxx.lanl.gov/abs/1110.1499}{{\tt arXiv:1110.1499}}].

\bibitem{Hirschi:2011pa}
V.~Hirschi, R.~Frederix, S.~Frixione, M.~V. Garzelli, F.~Maltoni, et~al., {\it
  {Automation of one-loop QCD corrections}},  {\em JHEP} {\bf 1105} (2011) 044,
  [\href{http://xxx.lanl.gov/abs/1103.0621}{{\tt arXiv:1103.0621}}].

\bibitem{Cullen:2011ac}
G.~Cullen, N.~Greiner, G.~Heinrich, G.~Luisoni, P.~Mastrolia, et~al., {\it
  {Automated One-Loop Calculations with GoSam}},  {\em Eur.Phys.J.} {\bf C72}
  (2012) 1889, [\href{http://xxx.lanl.gov/abs/1111.2034}{{\tt
  arXiv:1111.2034}}].

\bibitem{Badger:2012pg}
S.~Badger, B.~Biedermann, P.~Uwer, and V.~Yundin, {\it {Numerical evaluation of
  virtual corrections to multi-jet production in massless QCD}},  {\em
  Comput.Phys.Commun.} {\bf 184} (2013) 1981--1998,
  [\href{http://xxx.lanl.gov/abs/1209.0100}{{\tt arXiv:1209.0100}}].

\bibitem{Berger:2008sj}
C.~Berger, Z.~Bern, L.~Dixon, F.~Febres~Cordero, D.~Forde, et~al., {\it {An
  Automated Implementation of On-Shell Methods for One-Loop Amplitudes}},  {\em
  Phys.Rev.} {\bf D78} (2008) 036003,
  [\href{http://xxx.lanl.gov/abs/0803.4180}{{\tt arXiv:0803.4180}}].

\bibitem{Bredenstein:2010rs}
A.~Bredenstein, A.~Denner, S.~Dittmaier, and S.~Pozzorini, {\it {NLO QCD
  Corrections to Top Anti-Top Bottom Anti-Bottom Production at the LHC: 2. full
  hadronic results}},  {\em JHEP} {\bf 1003} (2010) 021,
  [\href{http://xxx.lanl.gov/abs/1001.4006}{{\tt arXiv:1001.4006}}].

\bibitem{Alioli:2010xd}
S.~Alioli, P.~Nason, C.~Oleari, and E.~Re, {\it {A general framework for
  implementing NLO calculations in shower Monte Carlo programs: the POWHEG
  BOX}},  {\em JHEP} {\bf 1006} (2010) 043,
  [\href{http://xxx.lanl.gov/abs/1002.2581}{{\tt arXiv:1002.2581}}].

\bibitem{Campbell:2011bn}
J.~M. Campbell, R.~K. Ellis, and C.~Williams, {\it {Vector boson pair
  production at the LHC}},  {\em JHEP} {\bf 1107} (2011) 018,
  [\href{http://xxx.lanl.gov/abs/1105.0020}{{\tt arXiv:1105.0020}}].

\bibitem{Arnold:2011wj}
K.~Arnold, J.~Bellm, G.~Bozzi, M.~Brieg, F.~Campanario, et~al., {\it {VBFNLO: A
  Parton Level Monte Carlo for Processes with Electroweak Bosons -- Manual for
  Version 2.5.0}},  \href{http://xxx.lanl.gov/abs/1107.4038}{{\tt
  arXiv:1107.4038}}.

\bibitem{Becker:2011vg}
S.~Becker, D.~Goetz, C.~Reuschle, C.~Schwan, and S.~Weinzierl, {\it {NLO
  results for five, six and seven jets in electron-positron annihilation}},
  {\em Phys.Rev.Lett.} {\bf 108} (2012) 032005,
  [\href{http://xxx.lanl.gov/abs/1111.1733}{{\tt arXiv:1111.1733}}].

\bibitem{Cascioli:2011va}
F.~Cascioli, P.~Maierhofer, and S.~Pozzorini, {\it {Scattering Amplitudes with
  Open Loops}},  {\em Phys.Rev.Lett.} {\bf 108} (2012) 111601,
  [\href{http://xxx.lanl.gov/abs/1111.5206}{{\tt arXiv:1111.5206}}].

\bibitem{Actis:2012qn}
S.~Actis, A.~Denner, L.~Hofer, A.~Scharf, and S.~Uccirati, {\it {Recursive
  generation of one-loop amplitudes in the Standard Model}},  {\em JHEP} {\bf
  1304} (2013) 037, [\href{http://xxx.lanl.gov/abs/1211.6316}{{\tt
  arXiv:1211.6316}}].

\bibitem{Ossola:2006us}
G.~Ossola, C.~G. Papadopoulos, and R.~Pittau, {\it {Reducing full one-loop
  amplitudes to scalar integrals at the integrand level}},  {\em Nucl.Phys.}
  {\bf B763} (2007) 147--169,
  [\href{http://xxx.lanl.gov/abs/hep-ph/0609007}{{\tt hep-ph/0609007}}].

\bibitem{Ossola:2007bb}
G.~Ossola, C.~G. Papadopoulos, and R.~Pittau, {\it {Numerical evaluation of
  six-photon amplitudes}},  {\em JHEP} {\bf 0707} (2007) 085,
  [\href{http://xxx.lanl.gov/abs/0704.1271}{{\tt arXiv:0704.1271}}].

\bibitem{Ossola:2007ax}
G.~Ossola, C.~G. Papadopoulos, and R.~Pittau, {\it {CutTools: a program
  implementing the OPP reduction method to compute one-loop amplitudes}},  {\em
  JHEP} {\bf 03} (2008) 042, [\href{http://xxx.lanl.gov/abs/0711.3596}{{\tt
  arXiv:0711.3596}}].

\bibitem{Mastrolia:2011pr}
P.~Mastrolia and G.~Ossola, {\it {On the Integrand-Reduction Method for
  Two-Loop Scattering Amplitudes}},  {\em JHEP} {\bf 1111} (2011) 014,
  [\href{http://xxx.lanl.gov/abs/1107.6041}{{\tt arXiv:1107.6041}}].

\bibitem{Badger:2012dp}
S.~Badger, H.~Frellesvig, and Y.~Zhang, {\it {Hepta-Cuts of Two-Loop Scattering
  Amplitudes}},  {\em JHEP} {\bf 1204} (2012) 055,
  [\href{http://xxx.lanl.gov/abs/1202.2019}{{\tt arXiv:1202.2019}}].

\bibitem{Zhang:2012ce}
Y.~Zhang, {\it {Integrand-Level Reduction of Loop Amplitudes by Computational
  Algebraic Geometry Methods}},  {\em JHEP} {\bf 1209} (2012) 042,
  [\href{http://xxx.lanl.gov/abs/1205.5707}{{\tt arXiv:1205.5707}}].

\bibitem{Mastrolia:2012an}
P.~Mastrolia, E.~Mirabella, G.~Ossola, and T.~Peraro, {\it {Scattering
  Amplitudes from Multivariate Polynomial Division}},  {\em Phys.Lett.} {\bf
  B718} (2012) 173--177, [\href{http://xxx.lanl.gov/abs/1205.7087}{{\tt
  arXiv:1205.7087}}].

\bibitem{Mastrolia:2013kca}
P.~Mastrolia, E.~Mirabella, G.~Ossola, and T.~Peraro, {\it {Multiloop Integrand
  Reduction for Dimensionally Regulated Amplitudes}},  {\em Phys.Lett.} {\bf
  B727} (2013) 532--535, [\href{http://xxx.lanl.gov/abs/1307.5832}{{\tt
  arXiv:1307.5832}}].

\bibitem{Giele:2008ve}
W.~T. Giele, Z.~Kunszt, and K.~Melnikov, {\it {Full one-loop amplitudes from
  tree amplitudes}},  {\em JHEP} {\bf 0804} (2008) 049,
  [\href{http://xxx.lanl.gov/abs/0801.2237}{{\tt arXiv:0801.2237}}].

\bibitem{Mastrolia:2010nb}
P.~Mastrolia, G.~Ossola, T.~Reiter, and F.~Tramontano, {\it {Scattering
  AMplitudes from Unitarity-based Reduction Algorithm at the Integrand-level}},
   {\em JHEP} {\bf 1008} (2010) 080,
  [\href{http://xxx.lanl.gov/abs/1006.0710}{{\tt arXiv:1006.0710}}].

\bibitem{Binoth:2005ff}
T.~Binoth, J.~P. Guillet, G.~Heinrich, E.~Pilon, and C.~Schubert, {\it {An
  Algebraic/numerical formalism for one-loop multi-leg amplitudes}},  {\em
  JHEP} {\bf 0510} (2005) 015,
  [\href{http://xxx.lanl.gov/abs/hep-ph/0504267}{{\tt hep-ph/0504267}}].

\bibitem{Heinrich:2010ax}
G.~Heinrich, G.~Ossola, T.~Reiter, and F.~Tramontano, {\it {Tensorial
  Reconstruction at the Integrand Level}},  {\em JHEP} {\bf 1010} (2010) 105,
  [\href{http://xxx.lanl.gov/abs/1008.2441}{{\tt arXiv:1008.2441}}].

\bibitem{Binoth:1999sp}
T.~Binoth, J.~Guillet, and G.~Heinrich, {\it {Reduction formalism for
  dimensionally regulated one loop N point integrals}},  {\em Nucl.Phys.} {\bf
  B572} (2000) 361--386, [\href{http://xxx.lanl.gov/abs/hep-ph/9911342}{{\tt
  hep-ph/9911342}}].

\bibitem{Reiter:2009kb}
T.~Reiter, {\it {Automated Evaluation of One-Loop Six-Point Processes for the
  LHC}},  \href{http://xxx.lanl.gov/abs/0903.0947}{{\tt arXiv:0903.0947}}.
  Ph.D. Thesis, The University of Edinburgh, 2008.

\bibitem{Cullen:2011kv}
G.~Cullen, J.~Guillet, G.~Heinrich, T.~Kleinschmidt, E.~Pilon, et~al., {\it
  {Golem95C: A library for one-loop integrals with complex masses}},  {\em
  Comput.Phys.Commun.} {\bf 182} (2011) 2276--2284,
  [\href{http://xxx.lanl.gov/abs/1101.5595}{{\tt arXiv:1101.5595}}].

\bibitem{Mastrolia:2008jb}
P.~Mastrolia, G.~Ossola, C.~Papadopoulos, and R.~Pittau, {\it {Optimizing the
  Reduction of One-Loop Amplitudes}},  {\em JHEP} {\bf 0806} (2008) 030,
  [\href{http://xxx.lanl.gov/abs/0803.3964}{{\tt arXiv:0803.3964}}].

\bibitem{Mastrolia:2012du}
P.~Mastrolia, E.~Mirabella, G.~Ossola, T.~Peraro, and H.~van Deurzen, {\it {The
  Integrand Reduction of One- and Two-Loop Scattering Amplitudes}},  {\em PoS}
  {\bf LL2012} (2012) 028, [\href{http://xxx.lanl.gov/abs/1209.5678}{{\tt
  arXiv:1209.5678}}].

\bibitem{vanDeurzen:2013pja}
H.~van Deurzen, {\it {Associated Higgs Production at NLO with GoSam}},  {\em
  Acta Phys.Polon.} {\bf B44} (2013), no.~11 2223--2230.

\bibitem{Peraro:2014cba}
T.~Peraro, {\it {Ninja: Automated Integrand Reduction via Laurent Expansion for
  One-Loop Amplitudes}},  \href{http://xxx.lanl.gov/abs/1403.1229}{{\tt
  arXiv:1403.1229}}.

\bibitem{vanDeurzen:2013saa}
H.~van Deurzen, G.~Luisoni, P.~Mastrolia, E.~Mirabella, G.~Ossola, et~al., {\it
  {Multi-leg One-loop Massive Amplitudes from Integrand Reduction via Laurent
  Expansion}},  \href{http://xxx.lanl.gov/abs/1312.6678}{{\tt
  arXiv:1312.6678}}.

\bibitem{Mastrolia:2012bu}
P.~Mastrolia, E.~Mirabella, and T.~Peraro, {\it {Integrand reduction of
  one-loop scattering amplitudes through Laurent series expansion}},  {\em
  JHEP} {\bf 1206} (2012) 095, [\href{http://xxx.lanl.gov/abs/1203.0291}{{\tt
  arXiv:1203.0291}}].

\bibitem{Greiner:2011mp}
N.~Greiner, A.~Guffanti, T.~Reiter, and J.~Reuter, {\it {NLO QCD corrections to
  the production of two bottom-antibottom pairs at the LHC}},  {\em
  Phys.Rev.Lett.} {\bf 107} (2011) 102002,
  [\href{http://xxx.lanl.gov/abs/1105.3624}{{\tt arXiv:1105.3624}}].

\bibitem{Greiner:2012im}
N.~Greiner, G.~Heinrich, P.~Mastrolia, G.~Ossola, T.~Reiter, et~al., {\it {NLO
  QCD corrections to the production of W+ W- plus two jets at the LHC}},  {\em
  Phys.Lett.} {\bf B713} (2012) 277--283,
  [\href{http://xxx.lanl.gov/abs/1202.6004}{{\tt arXiv:1202.6004}}].

\bibitem{vanDeurzen:2013rv}
H.~van Deurzen, N.~Greiner, G.~Luisoni, P.~Mastrolia, E.~Mirabella, et~al.,
  {\it {NLO QCD corrections to the production of Higgs plus two jets at the
  LHC}},  {\em Phys.Lett.} {\bf B721} (2013) 74--81,
  [\href{http://xxx.lanl.gov/abs/1301.0493}{{\tt arXiv:1301.0493}}].

\bibitem{Gehrmann:2013aga}
T.~Gehrmann, N.~Greiner, and G.~Heinrich, {\it {Photon isolation effects at NLO
  in $\gamma \gamma$ + jet final states in hadronic collisions}},  {\em JHEP}
  {\bf 1306} (2013) 058, [\href{http://xxx.lanl.gov/abs/1303.0824}{{\tt
  arXiv:1303.0824}}].

\bibitem{Luisoni:2013cuh}
G.~Luisoni, P.~Nason, C.~Oleari, and F.~Tramontano, {\it {$HW^{\pm}$/HZ + 0 and
  1 jet at NLO with the POWHEG BOX interfaced to GoSam and their merging within
  MiNLO}},  {\em JHEP} {\bf 1310} (2013) 083,
  [\href{http://xxx.lanl.gov/abs/1306.2542}{{\tt arXiv:1306.2542}}].

\bibitem{Hoeche:2013mua}
S.~Hoeche, J.~Huang, G.~Luisoni, M.~Schoenherr, and J.~Winter, {\it {Zero and
  one jet combined NLO analysis of the top quark forward-backward asymmetry}},
  {\em Phys.Rev.} {\bf D88} (2013) 014040,
  [\href{http://xxx.lanl.gov/abs/1306.2703}{{\tt arXiv:1306.2703}}].

\bibitem{Cullen:2013saa}
G.~Cullen, H.~van Deurzen, N.~Greiner, G.~Luisoni, P.~Mastrolia, et~al., {\it
  {NLO QCD corrections to Higgs boson production plus three jets in gluon
  fusion}},  {\em Phys.Rev.Lett.} {\bf 111} (2013) 131801,
  [\href{http://xxx.lanl.gov/abs/1307.4737}{{\tt arXiv:1307.4737}}].

\bibitem{vanDeurzen:2013xla}
H.~van Deurzen, G.~Luisoni, P.~Mastrolia, E.~Mirabella, G.~Ossola, et~al., {\it
  {NLO QCD corrections to Higgs boson production in association with a top
  quark pair and a jet}},  {\em Phys.Rev.Lett.} {\bf 111} (2013) 171801,
  [\href{http://xxx.lanl.gov/abs/1307.8437}{{\tt arXiv:1307.8437}}].

\bibitem{Gehrmann:2013bga}
T.~Gehrmann, N.~Greiner, and G.~Heinrich, {\it Precise qcd predictions for the
  production of a photon pair in association with two jets},  {\em Phys. Rev.
  Lett.} {\bf 111} (2013) 222002,
  [\href{http://xxx.lanl.gov/abs/1308.3660}{{\tt arXiv:1308.3660}}].

\bibitem{Dolan:2013rja}
M.~J. Dolan, C.~Englert, N.~Greiner, and M.~Spannowsky, {\it {Further on up the
  road: $hhjj$ production at the LHC}},  {\em Phys.Rev.Lett.} {\bf 112} (2014)
  101802, [\href{http://xxx.lanl.gov/abs/1310.1084}{{\tt arXiv:1310.1084}}].

\bibitem{Heinrich:2013qaa}
G.~Heinrich, A.~Maier, R.~Nisius, J.~Schlenk, and J.~Winter, {\it {NLO QCD
  corrections to WWbb production with leptonic decays in the light of top quark
  mass and asymmetry measurements}},
  \href{http://xxx.lanl.gov/abs/1312.6659}{{\tt arXiv:1312.6659}}.

\bibitem{Cullen:2012eh}
G.~Cullen, N.~Greiner, and G.~Heinrich, {\it {Susy-QCD corrections to
  neutralino pair production in association with a jet}},  {\em Eur.Phys.J.}
  {\bf C73} (2013) 2388, [\href{http://xxx.lanl.gov/abs/1212.5154}{{\tt
  arXiv:1212.5154}}].

\bibitem{Greiner:2013gca}
N.~Greiner, G.~Heinrich, J.~Reichel, and J.~F. von Soden-Fraunhofen, {\it {NLO
  QCD corrections to diphoton plus jet production through graviton exchange}},
  {\em JHEP} {\bf 1311} (2013) 028,
  [\href{http://xxx.lanl.gov/abs/1308.2194}{{\tt arXiv:1308.2194}}].

\bibitem{Binoth:2010xt}
T.~Binoth, F.~Boudjema, G.~Dissertori, A.~Lazopoulos, A.~Denner, et~al., {\it
  {A Proposal for a standard interface between Monte Carlo tools and one-loop
  programs}},  {\em Comput.Phys.Commun.} {\bf 181} (2010) 1612--1622,
  [\href{http://xxx.lanl.gov/abs/1001.1307}{{\tt arXiv:1001.1307}}].

\bibitem{Alioli:2013nda}
S.~Alioli, S.~Badger, J.~Bellm, B.~Biedermann, F.~Boudjema, et~al., {\it
  {Update of the Binoth Les Houches Accord for a standard interface between
  Monte Carlo tools and one-loop programs}},  {\em Comput.Phys.Commun.} {\bf
  185} (2014) 560--571, [\href{http://xxx.lanl.gov/abs/1308.3462}{{\tt
  arXiv:1308.3462}}].

\bibitem{gosamhome}
http://gosam.hepforge.org.

\bibitem{Nogueira:1991ex}
P.~Nogueira, {\it {Automatic Feynman graph generation}},  {\em J.Comput.Phys.}
  {\bf 105} (1993) 279--289.

\bibitem{Vermaseren:2000nd}
J.~Vermaseren, {\it {New features of FORM}},
  \href{http://xxx.lanl.gov/abs/math-ph/0010025}{{\tt math-ph/0010025}}.

\bibitem{Kuipers:2012rf}
J.~Kuipers, T.~Ueda, J.~Vermaseren, and J.~Vollinga, {\it {FORM version 4.0}},
  {\em Comput.Phys.Commun.} {\bf 184} (2013) 1453--1467,
  [\href{http://xxx.lanl.gov/abs/1203.6543}{{\tt arXiv:1203.6543}}].

\bibitem{Cullen:2010jv}
G.~Cullen, M.~Koch-Janusz, and T.~Reiter, {\it {Spinney: A Form Library for
  Helicity Spinors}},  {\em Comput.Phys.Commun.} {\bf 182} (2011) 2368--2387,
  [\href{http://xxx.lanl.gov/abs/1008.0803}{{\tt arXiv:1008.0803}}].

\bibitem{Reiter:2009ts}
T.~Reiter, {\it {Optimising Code Generation with haggies}},  {\em
  Comput.Phys.Commun.} {\bf 181} (2010) 1301--1331,
  [\href{http://xxx.lanl.gov/abs/0907.3714}{{\tt arXiv:0907.3714}}].

\bibitem{Ellis:2008ir}
R.~Ellis, W.~T. Giele, Z.~Kunszt, and K.~Melnikov, {\it {Masses, fermions and
  generalized $D$-dimensional unitarity}},  {\em Nucl.Phys.} {\bf B822} (2009)
  270--282, [\href{http://xxx.lanl.gov/abs/0806.3467}{{\tt arXiv:0806.3467}}].

\bibitem{Binoth:2008uq}
T.~Binoth, J.-P. Guillet, G.~Heinrich, E.~Pilon, and T.~Reiter, {\it {Golem95:
  A Numerical program to calculate one-loop tensor integrals with up to six
  external legs}},  {\em Comput.Phys.Commun.} {\bf 180} (2009) 2317--2330,
  [\href{http://xxx.lanl.gov/abs/0810.0992}{{\tt arXiv:0810.0992}}].

\bibitem{Guillet:2013msa}
J.~P. Guillet, G.~Heinrich, and J.~von Soden-Fraunhofen, {\it {Tools for NLO
  automation: extension of the golem95C integral library}},
  \href{http://xxx.lanl.gov/abs/1312.3887}{{\tt arXiv:1312.3887}}.

\bibitem{vanHameren:2010cp}
A.~van Hameren, {\it {OneLOop: For the evaluation of one-loop scalar
  functions}},  {\em Comput.Phys.Commun.} {\bf 182} (2011) 2427--2438,
  [\href{http://xxx.lanl.gov/abs/1007.4716}{{\tt arXiv:1007.4716}}].

\bibitem{vanOldenborgh:1990yc}
G.~van Oldenborgh, {\it {FF: A Package to evaluate one loop Feynman diagrams}},
   {\em Comput.Phys.Commun.} {\bf 66} (1991) 1--15.

\bibitem{Ellis:2007qk}
R.~K. Ellis and G.~Zanderighi, {\it {Scalar one-loop integrals for QCD}},  {\em
  JHEP} {\bf 02} (2008) 002, [\href{http://xxx.lanl.gov/abs/0712.1851}{{\tt
  arXiv:0712.1851}}].

\bibitem{Hahn:1998yk}
T.~Hahn and M.~Perez-Victoria, {\it {Automatized one loop calculations in
  four-dimensions and D-dimensions}},  {\em Comput.Phys.Commun.} {\bf 118}
  (1999) 153--165, [\href{http://xxx.lanl.gov/abs/hep-ph/9807565}{{\tt
  hep-ph/9807565}}].

\bibitem{Fleischer:2010sq}
J.~Fleischer and T.~Riemann, {\it {A Complete algebraic reduction of one-loop
  tensor Feynman integrals}},  {\em Phys.Rev.} {\bf D83} (2011) 073004,
  [\href{http://xxx.lanl.gov/abs/1009.4436}{{\tt arXiv:1009.4436}}].

\bibitem{Fleischer:2012et}
J.~Fleischer, T.~Riemann, and V.~Yundin, {\it {New developments in PJFry}},
  {\em PoS} {\bf LL2012} (2012) 020,
  [\href{http://xxx.lanl.gov/abs/1210.4095}{{\tt arXiv:1210.4095}}].

\bibitem{Actis:2013dfa}
S.~Actis, A.~Denner, L.~Hofer, A.~Scharf, and S.~Uccirati, {\it {EW and QCD
  One-Loop Amplitudes with RECOLA}},
  \href{http://xxx.lanl.gov/abs/1311.6662}{{\tt arXiv:1311.6662}}.

\bibitem{Catani:2000ef}
S.~Catani, S.~Dittmaier, and Z.~Trocsanyi, {\it {One loop singular behavior of
  QCD and SUSY QCD amplitudes with massive partons}},  {\em Phys.Lett.} {\bf
  B500} (2001) 149--160, [\href{http://xxx.lanl.gov/abs/hep-ph/0011222}{{\tt
  hep-ph/0011222}}].

\bibitem{Badger:2010nx}
S.~Badger, B.~Biedermann, and P.~Uwer, {\it {NGluon: A Package to Calculate
  One-loop Multi-gluon Amplitudes}},  {\em Comput.Phys.Commun.} {\bf 182}
  (2011) 1674--1692, [\href{http://xxx.lanl.gov/abs/1011.2900}{{\tt
  arXiv:1011.2900}}].

\bibitem{LesHouches2013}
Proceedings of the Les Houches 2013 workshop on Physics at TeV colliders, 2014.

\bibitem{Bellm:2013lba}
J.~Bellm, S.~Gieseke, D.~Grellscheid, A.~Papaefstathiou, S.~Pl{\"a}tzer,
  et~al., {\it {Herwig++ 2.7 Release Note}},
  \href{http://xxx.lanl.gov/abs/1310.6877}{{\tt arXiv:1310.6877}}.

\bibitem{Platzer:2011bc}
S.~Pl{\"a}tzer and S.~Gieseke, {\it {Dipole Showers and Automated NLO Matching
  in Herwig++}},  {\em Eur.Phys.J.} {\bf C72} (2012) 2187,
  [\href{http://xxx.lanl.gov/abs/1109.6256}{{\tt arXiv:1109.6256}}].

\bibitem{Denner:2005fg}
A.~Denner, S.~Dittmaier, M.~Roth, and L.~Wieders, {\it {Electroweak corrections
  to charged-current $e^+ e^- \to 4$ fermion processes: Technical details and
  further results}},  {\em Nucl.Phys.} {\bf B724} (2005) 247--294,
  [\href{http://xxx.lanl.gov/abs/hep-ph/0505042}{{\tt hep-ph/0505042}}].

\bibitem{Stelzer:1994ta}
T.~Stelzer and W.~Long, {\it {Automatic generation of tree level helicity
  amplitudes}},  {\em Comput.Phys.Commun.} {\bf 81} (1994) 357--371,
  [\href{http://xxx.lanl.gov/abs/hep-ph/9401258}{{\tt hep-ph/9401258}}].

\bibitem{Frederix:2008hu}
R.~Frederix, T.~Gehrmann, and N.~Greiner, {\it {Automation of the Dipole
  Subtraction Method in MadGraph/MadEvent}},  {\em JHEP} {\bf 0809} (2008) 122,
  [\href{http://xxx.lanl.gov/abs/0808.2128}{{\tt arXiv:0808.2128}}].

\bibitem{Frederix:2010cj}
R.~Frederix, T.~Gehrmann, and N.~Greiner, {\it {Integrated dipoles with
  MadDipole in the MadGraph framework}},  {\em JHEP} {\bf 1006} (2010) 086,
  [\href{http://xxx.lanl.gov/abs/1004.2905}{{\tt arXiv:1004.2905}}].

\bibitem{Alwall:2007st}
J.~Alwall, P.~Demin, S.~de~Visscher, R.~Frederix, M.~Herquet, et~al., {\it
  {MadGraph/MadEvent v4: The New Web Generation}},  {\em JHEP} {\bf 0709}
  (2007) 028, [\href{http://xxx.lanl.gov/abs/0706.2334}{{\tt
  arXiv:0706.2334}}].

\bibitem{SherpaDocu}
https://sherpa.hepforge.org/doc/SHERPA-MC-2.1.0.html.

\bibitem{gosamproc}
http://gosam.hepforge.org/proc/.

\bibitem{Degrande:2011ua}
C.~Degrande, C.~Duhr, B.~Fuks, D.~Grellscheid, O.~Mattelaer, et~al., {\it {UFO
  - The Universal FeynRules Output}},  {\em Comput.Phys.Commun.} {\bf 183}
  (2012) 1201--1214, [\href{http://xxx.lanl.gov/abs/1108.2040}{{\tt
  arXiv:1108.2040}}].

\bibitem{Christensen:2008py}
N.~D. Christensen and C.~Duhr, {\it {FeynRules - Feynman rules made easy}},
  {\em Comput.Phys.Commun.} {\bf 180} (2009) 1614--1641,
  [\href{http://xxx.lanl.gov/abs/0806.4194}{{\tt arXiv:0806.4194}}].

\bibitem{Alloul:2013bka}
A.~Alloul, N.~D. Christensen, C.~Degrande, C.~Duhr, and B.~Fuks, {\it
  {FeynRules 2.0 - A complete toolbox for tree-level phenomenology}},
  \href{http://xxx.lanl.gov/abs/1310.1921}{{\tt arXiv:1310.1921}}.

\bibitem{Semenov:2010qt}
A.~Semenov, {\it {LanHEP - a package for automatic generation of Feynman rules
  from the Lagrangian. Updated version 3.1}},
  \href{http://xxx.lanl.gov/abs/1005.1909}{{\tt arXiv:1005.1909}}.

\bibitem{Chetyrkin:1997un}
K.~Chetyrkin, B.~A. Kniehl, and M.~Steinhauser, {\it {Decoupling relations to O
  (alpha-s**3) and their connection to low-energy theorems}},  {\em Nucl.Phys.}
  {\bf B510} (1998) 61--87, [\href{http://xxx.lanl.gov/abs/hep-ph/9708255}{{\tt
  hep-ph/9708255}}].

\bibitem{ArkaniHamed:1998rs}
N.~Arkani-Hamed, S.~Dimopoulos, and G.~Dvali, {\it {The Hierarchy problem and
  new dimensions at a millimeter}},  {\em Phys.Lett.} {\bf B429} (1998)
  263--272, [\href{http://xxx.lanl.gov/abs/hep-ph/9803315}{{\tt
  hep-ph/9803315}}].

\bibitem{Antoniadis:1998ig}
I.~Antoniadis, N.~Arkani-Hamed, S.~Dimopoulos, and G.~Dvali, {\it {New
  dimensions at a millimeter to a Fermi and superstrings at a TeV}},  {\em
  Phys.Lett.} {\bf B436} (1998) 257--263,
  [\href{http://xxx.lanl.gov/abs/hep-ph/9804398}{{\tt hep-ph/9804398}}].

\bibitem{Kumar:2009nn}
M.~Kumar, P.~Mathews, V.~Ravindran, and A.~Tripathi, {\it {Direct photon pair
  production at the LHC to order $\alpha_s$ in TeV scale gravity models}},
  {\em Nucl.Phys.} {\bf B818} (2009) 28--51,
  [\href{http://xxx.lanl.gov/abs/0902.4894}{{\tt arXiv:0902.4894}}].

\bibitem{Stuart:1987tt}
R.~G. Stuart, {\it {Algebraic Reduction of One Loop Feynman Diagrams to Scalar
  Integrals}},  {\em Comput.Phys.Commun.} {\bf 48} (1988) 367--389.

\bibitem{Binoth:2006hk}
T.~Binoth, J.~P. Guillet, and G.~Heinrich, {\it {Algebraic evaluation of
  rational polynomials in one-loop amplitudes}},  {\em JHEP} {\bf 0702} (2007)
  013, [\href{http://xxx.lanl.gov/abs/hep-ph/0609054}{{\tt hep-ph/0609054}}].

\end{thebibliography}

\providecommand{\href}[2]{#2}\begingroup\raggedright\endgroup

\end{document}